\pdfoutput=1
\documentclass[final,1p,times]{elsarticle}

\usepackage{amsmath}
\usepackage{amssymb}
\usepackage{bbm}
\usepackage{graphicx,color}
\usepackage{tikz}
\usepackage{subcaption}
\graphicspath{{./}{figures/}}

\renewcommand{\vec}[1]{\textbf{#1}}
\newcommand{\mat}[1]{\textbf{#1}}
\newcommand{\tensor}[1]{\mathbbm{#1}}
\newcommand{\dd}{\mathrm{d}}
\newcommand{\diff}[1]{\dd#1}
\newcommand{\tr}{\operatorname{tr}}
\newcommand{\sym}{\operatorname{sym}}
\newcommand{\mom}[1]{\mu_#1}
\newcommand{\momd}[1]{\mu_#1}
\newcommand{\transpose}[1]{{#1}^\mathrm{T}}

\newcommand{\myunit}[2]{#1\,#2}
\newcommand{\isunit}[1]{{[#1]}}
\newcommand{\cf}{cf.\ }
\newcommand{\ie}{ie.\ }

\usepackage{booktabs}
\usepackage{tabularx}
\newcolumntype{P}[1]{>{\centering\arraybackslash\hspace{0pt}}p{#1}}
\usepackage{siunitx}

\usepackage{color}

\newcommand{\sdstep}[1]{\tikz[baseline=-1mm]{\draw circle (.25); \node[font=\footnotesize]{#1};}}
\newcommand{\sdexch}[1]{\tikz[baseline=-1mm]{\draw[shift={(0,-2.15000)}] (-0.250000,2.240000) -- (0.000000,2.365000) -- (0.250000,2.240000) -- (0.250000,1.990000) -- (-0.250000,1.990000) -- cycle; \node[font=\footnotesize]{#1};}}

\usepackage{enumitem}

\makeatletter
 \def\ps@pprintTitle{%
      \let\@oddhead\@empty
      \let\@evenhead\@empty
      \def\@oddfoot{\footnotesize\itshape
        Preprint submitted \hfill\today}%
      \let\@evenfoot\@oddfoot}
\makeatother

\begin{document}


\begin{frontmatter}
  \title{Enhanced conservation properties of Vlasov codes through coupling with conservative fluid models}

  \author{T. Trost}
  \author{S. Lautenbach}
  \author{R. Grauer}
  \ead{grauer@tp1.rub.de}
  \address{Institut f\"ur Theoretische Physik I, Ruhr-Universit\"at
    Bochum, 44801 Bochum, Germany}

  \begin{abstract}
    Many phenomena in collisionless plasma physics require a kinetic
    description. The evolution of the phase space density can be
    modeled by means of the Vlasov equation, which has to be solved
    numerically in most of the relevant cases. One of the problems
    that often arise in such simulations is the violation of important
    physical conservation laws. Numerical diffusion in phase space
    translates into unphysical heating, which can increase the overall
    energy significantly, depending on the time scale and the plasma
    regime. In this paper, a general and straightforward way of
    improving conservation properties of Vlasov schemes is presented
    that can potentially be applied to a variety of different
    codes. The basic idea is to use fluid models with good
    conservation properties for correcting kinetic models. The higher
    moments that are missing in the fluid models are provided by the
    kinetic codes, so that both kinetic and fluid codes compensate the
    weaknesses of each other in a closed feedback loop.
  \end{abstract}

\end{frontmatter}

\section{Introduction}

There is a variety of different plasma models with different strengths
and weaknesses. For collisionless plasmas, the Vlasov equation is
pretty accurate from a physical point of view, but it is typically
hard to solve analytically and expensive when it comes to numerical
simulations. Besides issues with reaching a reasonable resolution,
numerical Vlasov codes typically suffer from certain problems like
recurrence or artificial heating that are due to the discretization of
the phase space. The latter problem destroys energy conservation and
is thus highly unphysical. Fluid models only describe the evolution of
the (lowest) moments of the phase space density and are thus much
easier to handle than the full kinetic models, both analytically and
numerically. Here, the main problem lies in the fact that the
equations for the moments build an infinite hierarchy of dependent
expressions. While the hierarchy as a whole is still exact, only the
lowest moments can be examined in practice and approximating closures
determine the behavior of the final system of equations to a large
degree. Besides that, certain conservation properties are easy to guarantee in
numerical fluid codes.

At this point it becomes clear that kinetic and fluid models
complement each other: kinetic models offer higher moments and lack
certain conservation properties, while fluid models need information
on higher moments and have good conservation properties for the lower
moments. On the basis of this and taking into account that the
numerical costs of fluid models are negligible compared to those of
kinetic models, the basic idea is to use fluid codes alongside kinetic
codes in simulations and feed each of the codes information from the
other one in order to improve the overall results. The situation can
be imagined as follows: Two simulations of the same physical problem
with different models run parallely and mostly independent of each
other, but in each step the fluid code gets prescribed higher moments
as a closure from the kinetic code and the phase space density is
adapted to the moments given by the fluid model. The adaption or
\emph{fitting} of the phase space density is done through a linear
transformation of the velocity space that is a generalization of the
concept introduced in \cite{rieke:2015}. There, this fitting was used
for coupling kinetic and fluid codes in different regions, but the
idea stays the same.
The generalized version of the fitting procedure
can handle non-isotropic temperatures and may also be used for
coupling of fluid and kinetic codes between separated regions,
an area which is of major importance in fluid and plasma theory (see e.g.
\cite{arnold-giering:1997,degond-et-al:2010,dellacherie:2003,
      goudon-et-al:2013,klar-et-al:2000,tiwari-klar:1998,tiwari-et-al:2013,
      tallec-mallinger:1997,sugiyama-kusano:2007,daldorff-et-al:2014,
      markidis-et-al:2014}).

This paper is organized as follows: In \ref{sec:physical-models}, the
underlying physical equations and definitions are summarized in order
to clarify the general setup and for later reference. The numerical
schemes that are used for solving these equations are briefly
described in \ref{sec:numerical-schemes}. After that, the fitting
procedure that is used for correcting the phase space density on the
basis of its moments is presented in \ref{sec:fitting}. Various tests
of the correction procedure are discussed in
\ref{sec:benchmarks}. Finally, a summary of the results and an
outlook are given in \ref{sec:conclusion}.
\section{Physical models}\label{sec:physical-models}
The focus of the present paper lies on classical, non-relativistic,
collisionless electron/ion plasmas. Such plasmas can be described in
terms of the phase space densities $f_s(\vec{x},\vec{v},t)$, where the
subscript $s$ may be replaced by $e$ for electrons or $i$ for
ions. Each $f_s$ is governed by the non-relativistic Vlasov equation
\begin{equation}
  \label{eq:vlasov}
  \partial_t f_s + \vec{v}\cdot \nabla_{\vec{x}}f_s + \frac{q_s}{m_s}\big(\vec{E} + \vec{v} \times \vec{B} \big)\cdot \nabla_{\vec{v}}f_s = 0,
\end{equation}
where $q_s$ is the charge of a particle of species $s$ and $m_s$ is
its charge.  A closure for the electric field $\vec E$ and the
magnetic field $\vec B$ is given by Maxwell's equations:
\begin{subequations}\label{eq:maxwells-equations}
  \begin{align}
    \nabla \cdot \vec E &= \frac{\rho}{\varepsilon_0} \\
    \nabla \cdot \vec B &= 0 \\
    \partial_t \vec B &= - \nabla \times \vec E \label{eq:faradays-law} \\
    \partial_t \vec E &= c^2\left(\nabla\times\vec B - \mu_0 \vec J\right) \label{eq:amperes-law}
  \end{align}
\end{subequations}
Charge density $\rho$ and current density $\vec J$, however, depend on $f_e$ and $f_i$. They are defined as:
\begin{subequations}
  \begin{align}
    \rho &:= \sum_s q_s \int f_s\diff{\vec{v}}, \label{eq:charge-f} \\
    \vec J &:= \sum_s q_s \int \vec v f_s \diff{\vec{v}}. \label{eq:current-f}
  \end{align}
\end{subequations}
Even though \eqref{eq:vlasov} offers a comprehensive description of
the plasmas under consideration, solving it may be very challenging
and the full phase space density is not always of interest, anyway.
Thus, it is possible to consider the moments $\mu_{n,s}$ of $f_s$, instead,
\begin{equation}\label{eq:def-nth-moment}
  \mu_{n,s} := \int \vec{v}^n f_s \diff{\vec v}.
\end{equation}
The lowest moments that are relevant for the subsequent discussion are given separate names:
\begin{subequations}
  \begin{flalign}
    \label{eq:mass-density}
    &\text{particle density:} & n_s &:= \mu_{0,s} = \int f_s \diff v &\\
    \label{eq:momentum-density}
    &\text{bulk velocity:} & \vec u_s &:= \frac{\mu_{1,s}}{\mu_{0,s}}  = \frac{1}{n_s}\int \vec v f_s \diff v &\\
    \label{eq:energy-tensor}
    &\text{energy density tensor:} & \tensor{E}_s &:= m_s \mu_{2,s} = m_s\int \vec v^2 f_s \diff v &\\
    \label{eq:heat-flux}
    &\text{heat flux tensor:} & \tensor{Q}_s &:= m_s \mu_{3,s} = m_s\int \vec v^3 f_s \diff v &
  \end{flalign}
\end{subequations}
From that, additional quantities can be derived, as for example the pressure tensor
\begin{equation}
  \tensor{P}_s = \tensor{E}_s - m_s n_s \vec{u}_s\vec{u}_s,
\end{equation}
the temperature tensor
\begin{equation} \label{eq:temperature-tensor}
  \tensor{T}_s = \frac{1}{k_\mathrm{B} n_s}\tensor{P}_s
\end{equation}
or the scalar pressure $p_s = \frac 1 3 \tr \tensor{P}_s$ and the
scalar temperature $T_s = \frac 1 3 \tr \tensor{T}_s$.  Furthermore,
the scalar energy density
\begin{equation}
  \mathcal{E}_s = \frac 1 2 \tr \tensor{E}_s
\end{equation}
and the vector heat flux
\begin{equation}
  \vec{Q}_s = \begin{pmatrix}\tensor{Q}_{111,s}+\tensor{Q}_{122,s}+\tensor{Q}_{133,s}\\\tensor{Q}_{112,s}+\tensor{Q}_{222,s}+\tensor{Q}_{233,s}\\\tensor{Q}_{113,s}+\tensor{Q}_{223,s}+\tensor{Q}_{333,s}\end{pmatrix}
\end{equation}
can be introduced. Based on \eqref{eq:vlasov}, the first moments are then subject to the following set of equations:
\begin{subequations}
  \begin{align}
    \partial_t n_s =& - \nabla \cdot (n_s\vec{u}_s) \label{eq:continuity-3d} \\
    \partial_t (m_sn_s\vec{u}_s) =& - \nabla\cdot\tensor{E}_s + q_s \big( n_s\vec{E} + n_s\vec{u}_s \times \vec B \big) \label{eq:momentum-3d} \\
    \partial_t \tensor{E}_s =& -\nabla\cdot\tensor{Q}_s + 2 q_s \sym\left(n_s\vec{u}_s \vec E + \frac 1 {m_s}\tensor{E}_s\times \vec B\right) \label{eq:energy-3d}
  \end{align}
\end{subequations}
The scalar energy density evolves in time according to:
\begin{equation}
  \partial_t\mathcal{E}_s = -\nabla\cdot\vec{Q}_s - \nabla\cdot\left(\frac{5} 2 p_s\vec{u}_s - \frac 1 2 m_sn_s(\vec{u}_s\cdot\vec{u}_s)\vec{u}_s \right) + q_sn_s\vec{u}_s\cdot \vec E
\end{equation}
While these fluid equations are exact, the hierarchy is not closed and
would in this case require some kind of closure for $\tensor{Q}_s$
respectively $\vec{Q}_s$. For the pure fluid runs that are discussed
below, where this closure is not given through information from the
Vlasov solver, the equation for the scalar energy density is closed by
the assumption of adiabaticity, $\nabla\cdot\vec{Q}_s \equiv 0$, and
the full ten-moment model is closed by means of the ansatz
\begin{equation}\label{eq:closure-ten-moment}
    \nabla \cdot \tensor{Q}_s = v_{\mathrm{th},s}|k_0|(\tensor{P}_s - p_s\tensor 1)
\end{equation}
from \citet{wang-et-al:2015} (see also \cite{johnson:2013}). $k_0$ is a constant that has to be
prescribed for each species and $v_{\mathrm{th},s} = \sqrt{\frac{k_\mathrm{B}T_s}{m_s}}$ is the thermal velocity.

\section{Used numerical schemes}\label{sec:numerical-schemes}
The numerical methods used here have already been described in detail
in \cite{rieke:2015}. Because of that, only a brief description is
given in the following. Details with respect to the interaction of the
schemes are given below in \ref{ssec:interplay}.
\subsection{Vlasov solver}
The Vlasov equation is split into two separate equations by means of
Strang splitting; one with the gradient in physical space and one with
the gradient in velocity space. While the first of these can be split
into three one-dimensional equations without further approximations,
the so-called backsubstitution method \cite{schmitz1,schmitz2} (see also \cite{umeda-fukazawa} for applications) is used
for splitting the second one into three separate equations. The
resulting set of six one-dimensional equations is then solved with a
semi-Lagrangian scheme which is based on moving the probability
density along the characteristics of the differential equation. The
required interpolation is done by means of the positive
flux-conservative finite volume scheme introduced by
\citet{filbet2001conservative}. The resulting scheme is of second
order in space and time.
\subsection{Fluid solver}\label{sec:fluid-solver}
All fluid equations have the form of conservation equations and can
thus be solved with the same numerical scheme. Here, the CWENO scheme
\cite{kur2000}, a third order finite volume scheme, is used for
discretization in space. The resulting semi-discrete equations are
then solved with a third order strong-stability-preserving Runge-Kutta
scheme \cite{shu88}.
\subsection{Maxwell solver}
The electromagnetic fields are discretized on a Yee grid
\cite{yee1966} and evolved through the second order FDTD method
\cite{taf1975}.

\section{Fitting of the phase space density}\label{sec:fitting}
For a given time and at a given point in physical space, the phase
space density $f(\vec x, \vec v, t)$ that was defined in
\ref{sec:physical-models} is a simple function of the velocity
coordinates, $f(\vec v)$. An affine linear transformation of this
three-dimensional velocity space leads to a change of the moments of
$f$, which can be used for adapting $f$ to a given set of precribed
moments. The advantage of a linear transformation is that no new
features are introduced in the phase space density; the basic shape of
$f$ remains the same. Thus, a transformation that is close to identity
can be seen as a small correction of $f$, if the new set of moments is
considered more reliable than the original moments of $f$. Here, the
process of adapting $f$ will be denoted as \emph{fitting}.

In \cite{rieke:2015}, a fitting procedure on the basis of an
artificial advection step in velocity space was presented. While it
was particularly easy to implement, its main drawback was that it
lacked support for adapting $f$ to a non-isotropic temperature. In the
following, a more general affine linear fitting procedure is described
that is based on a slightly different ansatz but that comes with full
support for fitting up to the full second moment. The old procedure
turns out to be a special case of the new one. From a numerical point
of view, the Vlasov solver described above in
\ref{sec:numerical-schemes} can also be used for applying the fitting
transformation as it is basically an advection, too. Because of this,
the main objective is to find the characteristics of the respective
advection step. Besides that, the new method requires the numerical
solution of the resulting equations. After these have been discussed,
the overall time scheme for the fitting correction in the simulation
is given.

\subsection{Mathematical formulation}
A three-dimensional distribution function $f$ is to be adapted to a
given density, a given velocity and a given (full) temperature tensor.
Instead of formulating the problem in terms of a differential equation
(c.f. \cite{rieke:2015}), it is possible to think in terms of the
final goal, the characteristics, and make a direct ansatz for the
mapping of a point with coordinates $\vec{v}^\text{old}$ to its new position
$\vec{v}^\text{new}$:
\begin{equation}\label{eq:ansatz-10-mom-fit}
\vec{v}^\text{old} \mapsto \vec{b}^\prime + \mat{A}^{-1}(\vec{v}^\text{old} - \vec b) = \vec{v}^\text{new}
\end{equation}
Here, $\mat A$ is a real invertible $3\times 3$ matrix and $\vec
b,\;\vec{b}^\prime\in\mathbb{R}^3$. These parameters may be chosen
freely and in a manner that a distribution $f$ in the respective space is
transformed in such a way that its new moments after the
transformation correspond to given values. One of the $\vec b$'s is theoretically superfluous (for
getting this form of equation) and through the parameters there are
more degrees of freedom than necessary for adapting ten moments, but
it turns out that this ansatz is advantageous for solving the
problem.

In this and the next sections, the problem of choosing the parameters $\mat A$,
$\vec b$, and $\vec{b}^\prime$ in the right way is discussed. All
quantities related to the original distribution $f^\mathrm{old}$ are
marked with a superscript ``old'', while all the target
quantities are indicated via a superscript ``new''.

Under the transformation \eqref{eq:ansatz-10-mom-fit}, the raw
moments, as they are defined in \eqref{eq:def-nth-moment}, transform
in the following way:
\begin{align}
  \mom{n}^\mathrm{new} &= \int \vec{v}^nf^\mathrm{new}(\vec v)\diff{\vec v} = \int \vec{v}^nf^\mathrm{old}\left(\mat{A}(\vec v - \vec{b}^\prime) + \vec{b}\right)|\det \mat A| \diff{\vec v} \nonumber \\
  &= \int \left(\mat{A}^{-1}(\vec v - \vec b) + \vec{b}^\prime\right)^nf^\mathrm{old}(\vec v) \diff{\vec v}
\end{align}
Under this transformation the macroscopic density does not change:
\begin{equation}
 \int f^\mathrm{new}(\vec v) \diff{\vec v}  = \int f^\mathrm{old}(\vec v) \diff{\vec v} = \momd{0}^\mathrm{old} =: \momd{0}
\end{equation}
Adaption of the density is carried out at the beginning by multiplying
the distribution with the factor
$\momd{0}^\mathrm{new}/\momd{0}^\mathrm{old}$. The new first moment is
\begin{align}
  \mom{1}^\mathrm{new} &= \int \left(\mat{A}^{-1}(\vec v - \vec b) + \vec{b}^\prime\right)f^\mathrm{old}(\vec v) \diff{\vec v} \nonumber \\
  &= \mat{A}^{-1}\int \vec v f^\mathrm{old}(\vec v)\diff{\vec v} - (\mat{A}^{-1}\vec b - \vec{b}^\prime)\int f^\mathrm{old}(\vec v)\diff{\vec v} \nonumber \\
  &= \mat{A}^{-1}\mom{1}^\mathrm{old} - (\mat{A}^{-1}\vec b - \vec{b}^\prime)\momd{0}.
\end{align}
With the ansatz
\begin{subequations}\label{eq:10-mom-fit-bs}
  \begin{align}
    \vec b &= \frac{\mom{1}^\mathrm{old}}{\momd{0}} \\
    \vec{b}^\prime &= \frac{\mom{1}^\mathrm{new}}{\momd{0}}
  \end{align}
\end{subequations}
these equations are always fulfilled. Using \eqref{eq:10-mom-fit-bs}, the new second moment assumes the form
\begin{align}
  \mom{2}^\mathrm{new} &= \int \left(\mat{A}^{-1}(\vec v - \vec b) + \vec{b}^\prime\right)\left(\mat{A}^{-1}(\vec v - \vec b) + \vec{b}^\prime\right)f^\mathrm{old}(\vec v)\diff{\vec v} \nonumber \\
&= \mat{A}^{-1}\mom{2}^\mathrm{old}\transpose{\left(\mat{A}^{-1}\right)} - \frac{1}{\momd{0}}\mat{A}^{-1}\mom{1}^\mathrm{old}\mom{1}^\mathrm{old}\transpose{\left(\mat{A}^{-1}\right)} + \frac{1}{\momd{0}}\mom{1}^\mathrm{new}\mom{1}^\mathrm{new}
\end{align}
With $\tensor T = \mom{2}/\momd{0} - \mom{1}\mom{1}/\momd{0}^2$ this is equivalent to
\begin{equation}\label{eq:10-mom-fit-a}
  \tensor{T}^\mathrm{old} =  \mat A \tensor{T}^\mathrm{new} \transpose{\mat{A}},
\end{equation}
which is the final equation that has to be solved for getting a
suitable $\mat A$. This relation contains three equations for nine
parameters and is thus underdetermined and ill-posed. Nevertheless,
one is only interested in finding \emph{some} solution that behaves
well. A suitable additional condition is that $\mat A$ is as close to
unity as possible, which makes the overall correction only as large as
necessary.

\subsection{Numerical approach for finding the fitting matrix}\label{ssec:fitting-regularization}
The numerical approach for solving the problem \eqref{eq:10-mom-fit-a}
is associated with \emph{Tikhonov regularization}
\cite{tikhonov:1963}. The ill-posed problem is recast into a
minimization problem for the quantity:
\begin{equation}\label{eq:10-mom-fit-tikhonov}
  Z_A := \Vert\mat A\tensor{T}^\mathrm{new}\transpose{\mat A} - \tensor{T}^\mathrm{old}\Vert_2^2 + \lambda\Vert\mat A - \tensor{1}\Vert_2^2
\end{equation}
The first part of $Z_A$ expresses the actual equation that is to be
solved while the second part is a penalty term that sanctions large
deviations of $\mat A$ from unity. This way, small corrections are
preferred over large ones. The real parameter $\lambda\in\mathbb{R}_0^+$ regulates the
relative influence of the two terms.

Due to the product of $\mat A$ and $\transpose{\mat A}$,
\eqref{eq:10-mom-fit-tikhonov} is a forth order polynomial in the nine
components of $\mat A$ and the minimization problem is hard to solve
analytically. In the next paragraphs, two methods for finding an $\mat
A$ that minimizes $Z_A$ are presented.

\subsubsection{Gradient descent}
Contrary to the analytical approach, the numerical solution is
straightforward. A simple gradient descent (a method that was
originally proposed by \citet{cauchy:1847} and can now be found in any
textbook with a chapter on numerical optimization, e.g.
\cite{press:2007}), where $\mat A$ is initialized with the identity
matrix, already leads to reasonable results. The components of the
gradient of $Z_A$ are somewhat lengthy, but there is no fundamental
problem in calculating them:
\begin{equation}
  \frac{\partial Z_A}{\partial A_{mn}} = 4\sum_i\left(\sum_{k,\ell}A_{i\ell}A_{mk}T^\mathrm{new}_{k\ell}-T^\mathrm{old}_{mi}\right)\left(\sum_kA_{ik}T^\mathrm{new}_{kn}\right) + 2\lambda\left(A_{mn}-\delta_{mn}\right)
\end{equation}

\subsubsection{Analytical solution of simplified problem}
Introduce the new quantity
\begin{equation}
  \tilde{\mat{A}} := \mat A - \tensor{1}.
\end{equation}
that represents the deviation of $\mat A$ from unity. With that,
\eqref{eq:10-mom-fit-tikhonov} becomes
\begin{equation}\label{eq:10-mom-fit-tikhonov-alt}
  Z_A = \Vert \tensor{T}^\mathrm{new} - \tensor{T}^\mathrm{old} + \tilde{\mat{A}}\tensor{T}^\mathrm{new} + \tensor{T}^\mathrm{new}\transpose{\tilde{\mat{A}}} + \tilde{\mat{A}}\transpose{\tilde{\mat{A}}}\Vert_2^2 + \lambda\Vert\tilde{\mat{A}}\Vert_2^2.
\end{equation}
In actual simulations, all components of $\tilde{\mat{A}}$ are much
smaller than one because the fitting is merely a correction. If
$\tensor{T}^\mathrm{new}$ has components of order unity, which is
usually the case here, the quadratic term
$\tilde{\mat{A}}\transpose{\tilde{\mat{A}}}$ in
\eqref{eq:10-mom-fit-tikhonov-alt} can be neglected compared to the
others summands. The resulting expression
\begin{equation}\label{eq:10-mom-fit-tikhonov-approx}
  \tilde{Z}_A := \Vert \tensor{T}^\mathrm{new} - \tensor{T}^\mathrm{old} + \tilde{\mat{A}}\tensor{T}^\mathrm{new} + \tensor{T}^\mathrm{new}\transpose{\tilde{\mat{A}}}\Vert_2^2 + \lambda\Vert\tilde{\mat{A}}\Vert_2^2 \approx Z_A
\end{equation}
is much easier to handle than \eqref{eq:10-mom-fit-tikhonov} and can
be minimized analytically by searching for points with a vanishing
gradient. Starting with the explicit form of $\tilde{Z}_A$
\begin{equation}\label{eq:10-mom-fit-tikhonov-alt-explicit}
  \tilde{Z}_A = \sum_{i,j}\left\{\left(T^\mathrm{new}_{ij} - T^\mathrm{old}_{ij} + \sum_k\left(\tilde{A}_{ik}T^\mathrm{new}_{kj} + \tilde{A}_{jk}T^\mathrm{new}_{ki}\right)\right)^2 + \lambda \tilde{A}_{ij}^2\right\},
\end{equation}
the derivative of $\tilde{Z}_A$ with respect to the components of
$\tilde{\mat{A}}$ can readily be calculated:
\begin{equation}\label{eq:10-mom-fit-tikhonov-alt-grad}
  \frac{\partial \tilde{Z}_A}{\partial \tilde{A}_{mn}} = 2\sum_{i,j}\left(T^\mathrm{new}_{ij} - T^\mathrm{old}_{ij} + \sum_k\left(\tilde{A}_{ik}T^\mathrm{new}_{kj} + \tilde{A}_{jk}T^\mathrm{new}_{ki}\right)\right)\left(\delta_{im}T^\mathrm{new}_{nj}+\delta_{jm}T^\mathrm{new}_{in}\right) + \lambda\tilde{A}_{mn}
\end{equation}
\eqref{eq:10-mom-fit-tikhonov-alt-explicit} is a quadratic expression
in the nine components of $\tilde{\mat{A}}$ and thus smooth and
well-defined in the complete space $\mathbb{R}^9$ of these
components. If $\lambda$ is large enough there must be a unique global
minimum of this function. This optimum $\tilde{\mat{A}}$ is then the
point where all derivatives \eqref{eq:10-mom-fit-tikhonov-alt-grad}
vanish and can be found by solving the associated linear system of
equations. The coefficients of the respective matrix are
\begin{align}
  \frac{\partial^2\tilde{Z}_A}{\partial \tilde{A}_{mn}\partial \tilde{A}_{\mu\nu}} =& 2\sum_{i,j}\left(\delta_{im}T^\mathrm{new}_{nj}+\delta_{jm}T^\mathrm{new}_{in}\right)\left(\delta_{i\mu}T^\mathrm{new}_{\nu j}+\delta_{j\mu}T^\mathrm{new}_{i\nu}\right) + 2\lambda\delta_{m\mu}\delta_{n\nu} \nonumber \\
  =& 4\delta_{m\mu}\sum_iT^\mathrm{new}_{i\nu}T^\mathrm{new}_{in}+4T^\mathrm{new}_{m \nu}T^\mathrm{new}_{n\mu}+ 2\lambda\delta_{m\mu}\delta_{n\nu}
\end{align}
and the inhomogeneity in the equation for $\tilde{A}_{mn}$, i.e. the
part of \eqref{eq:10-mom-fit-tikhonov-alt-grad} that does not contain
any components of $\tilde{\mat{A}}$ as a factor, is given by
\begin{equation}
  2\sum_{i,j}\left(T^\mathrm{old}_{ij} - T^\mathrm{new}_{ij}\right)\left(\delta_{im}T^\mathrm{new}_{nj}+\delta_{jm}T^\mathrm{new}_{in}\right) = 4\sum_iT^\mathrm{new}_{in}\left(T^\mathrm{old}_{im}-T^\mathrm{new}_{im}\right).
\end{equation}

\subsection{Backsubstitution method}
As it was mentioned above, one central advantage of the way in which
fitting is approached here is that it can directly be plugged into the
existing PFC scheme used for the Vlasov solver. For doing so in an
effective way, eqn.~\eqref{eq:ansatz-10-mom-fit} has to be recast in a way
appropriate for the backsubstition method \cite{schmitz1,leslie-purser:1995}. This basically means that in
the new form, the fitting is split into three one-dimensional steps
along the three coordinate axes. With the short form
\begin{subequations}
  \begin{align}
    \vec{w}^+ &= \vec{v}^+ - {\vec b}^\prime, \\
    \vec{w}^- &= \vec{v}^- - {\vec b},
  \end{align}
\end{subequations}
\eqref{eq:ansatz-10-mom-fit} can be reformulated in the form:
\begin{subequations}
\begin{align}
  w_x^- &=  \frac{1}{(A_{yz}A_{zy} - A_{yy}A_{zz})}\Big\{
	      (A_{xy}A_{zz}-A_{xz}A_{zy})(A_{yx}w_x^+ - w_y^-)
               \nonumber \\
	      &+ (A_{xz}A_{yy} - A_{xy}A_{yz})(A_{zx}w_x^+ - w_z^-)\Big\} + A_{xx}w_x^+ \\
  w_y^- &= A_{yx}w_x^+ + \frac{A_{yz}}{A_{zz}}(w_z^- - A_{zx}w_x^+ - A_{zy}w_y^+) + A_{yy}w_y^+ \\
  w_z^- &= A_{zx}w_x^+ + A_{zy}w_y^+ + A_{zz}w_z^+
\end{align}
\end{subequations}
See also \cite{rieke:2015} for more details on how to use this in a program.
\subsection{Interplay of models}\label{ssec:interplay}
\begin{figure}
  \includegraphics{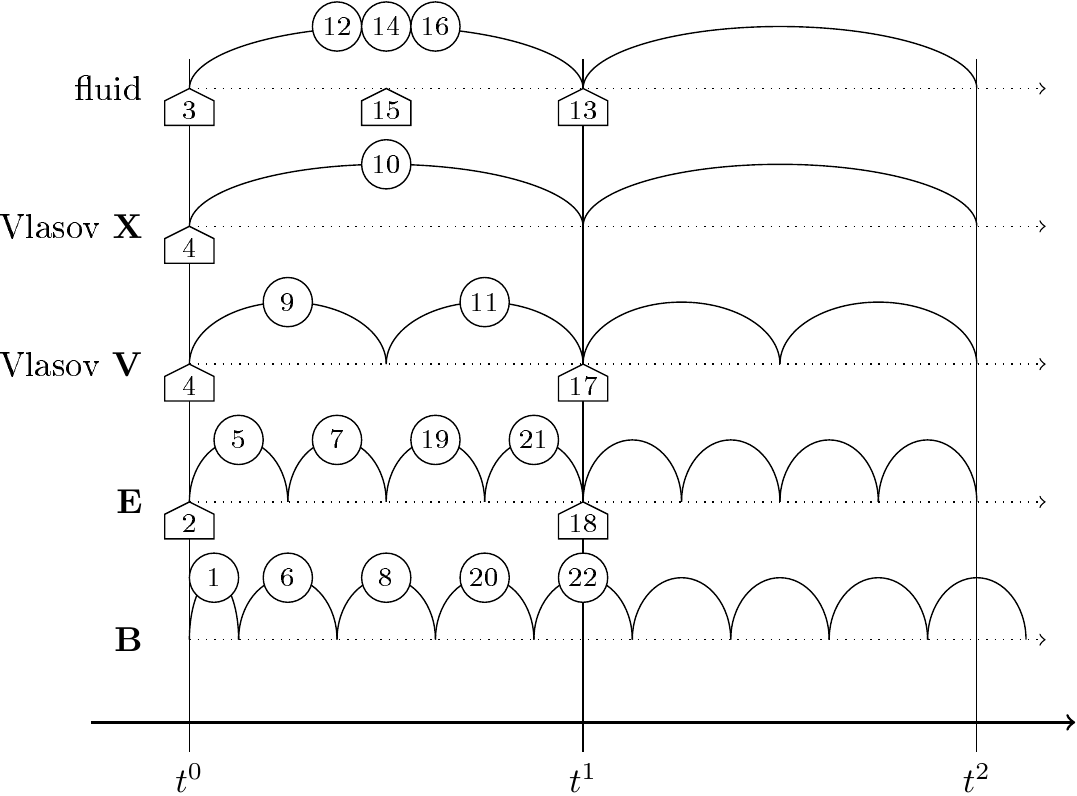}
  \begin{description}[align=right, labelwidth=1.5cm, font=\normalfont, leftmargin=!]
    \itemsep-1mm
  \item[\sdstep{1}] Start leap frog scheme with half a step of magnetic field (not repeated).
  \item[\sdexch{2}] Calculate initial currents and send them to Maxwell solver (not repeated).
  \item[\sdexch{3} \sdexch{4}] Distribute electromagnetic fields.
  \item[\sdstep{5} -- \sdstep{8}] Advance electromagnetic fields to
    time $t_0 + \frac 1 2 \Delta t$ respectively half a substep
    further through subcycling.
  \item[\sdstep{9} -- \sdstep{11}] Complete step of Vlasov solver.
  \item[\sdstep{12}] Euler step of fluid solver, so that fluid quantities live at time $t_1$.
  \item[\sdexch{13}] Set higher moments of fluid from phase space
    density that already lives at $t_1$.
  \item[\sdstep{14}] Second Runge-Kutta substep of fluid solver.
  \item[\sdexch{15}] The higher moments at time
    $t_0 + \frac 1 2 \Delta t$ are gained through interpolation. The
    electromagnetic fields live at the correct time and are directly
    copied.
  \item[\sdstep{16}] Last Runge-Kutta substep of fluid solver.
  \item[\sdexch{17}] Correct phase space density on basis of fluid quantities.
  \item[\sdexch{18}] Calculate currents at $t_1$ and send them to Maxwell solver.
  \item[\sdstep{19} -- \sdstep{22}] Advance electromagnetic fields to
    time $t_1$ respectively half a substep further. The value of $\vec
    B$ at time $t_1$ can be approximated through interpolation.
  \end{description}
  \caption[Time layout for coupling of electromagnetic models]{Stepping scheme for coupling of three-dimensional fluid-, Vlasov and Maxwell equations.}
  \label{fig:time-layout-coupling-2d3v}
\end{figure}
In the explicit schemes that are used here, each physical quantity
always lives at a specific individual time, depending on how far the
respective scheme has proceeded. During the simulation, these times
may differ between the various quantities and the interstations may be
different for each scheme. It is important that such times fit
together when the respective schemes communicate with each other.If
these time issues are not addressed, a variety of instabilities can
occur, depending on the respective situation.  In order to adapt them
to each other the schemes are subdivided into atomic substeps, \ie
smallest units of time stepping after that a stop leaves the data in a
valid state with a well-defined time. These substeps are carried out
in such an order that all the data has the appropriate time if
communication has to take place. For that to be possible it may be
necessary to adapt the numerical schemes slightly, for example by
choosing a suitable Runge-Kutta method or artificially subdividing
parts of the update procedure for obtaining intermediate results. In
some cases, interpolation may be necessary for getting appropriate
data.

The time layout of the coupling between the three-dimensional Vlasov
solver, fluid models and Maxwell's equations is shown and described in
\ref{fig:time-layout-coupling-2d3v}. Because the time scale given by the
speed of light normally exceeds other time scales in the simulation by
far, sub-cycling for the electromagnetic may be used to some degree in
order to avoid unnecessarily many expensive Vlasov steps.
\section{Benchmarks}\label{sec:benchmarks}
In order to test the correction scheme, several benchmark problems are
considered. First, a preliminary examination of the basic idea is
conducted: Different two-dimensional distributions are rotated with
constant angular velocity by means of the advection scheme of the
Vlasov solver described above\footnote{This roughly corresponds to a
  setup with a prescribed constant magnetic field and a vanishing
  electric field.}. In a discretized simulation like that, the
gyration leads to a purely artificial smearing of the
distribution. The mellowing of this smearing by means of correction
fitting is examined quantitatively. After this preliminary
investigation, full simulations of a fast magnetosonic wave with and
without correction fitting are considered as an example of a linear
phenomenon with anisotropic pressure.  The final benchmark for
assessing correction fitting in a fully non-linear regime is a
simulation of the GEM challenge with different numerical models, with
and without correction fitting.

\subsection{Fundamental test}
As a first test case, the evolution of a Maxwellian-shaped, slightly decentered
bulb with an initially rotational velocity distribution is observed in
two-dimensional $(v_x,v_y)$-space.
\begin{figure}
  \centering
  \subcaptionbox{density}{\includegraphics[width=.32\textwidth]{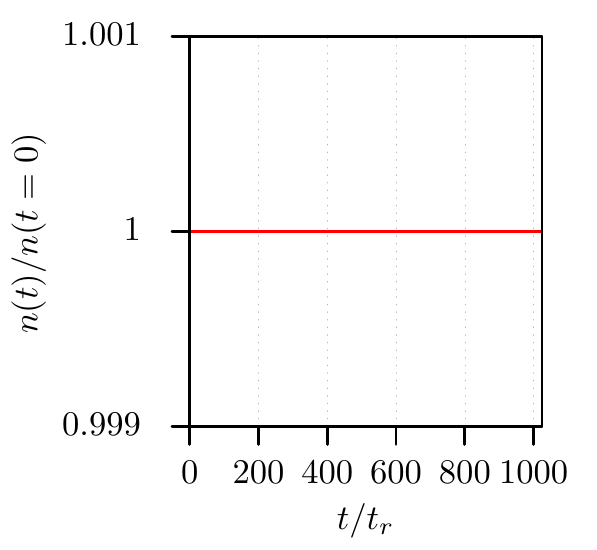}}
  \subcaptionbox{temperature}{\includegraphics[width=.32\textwidth]{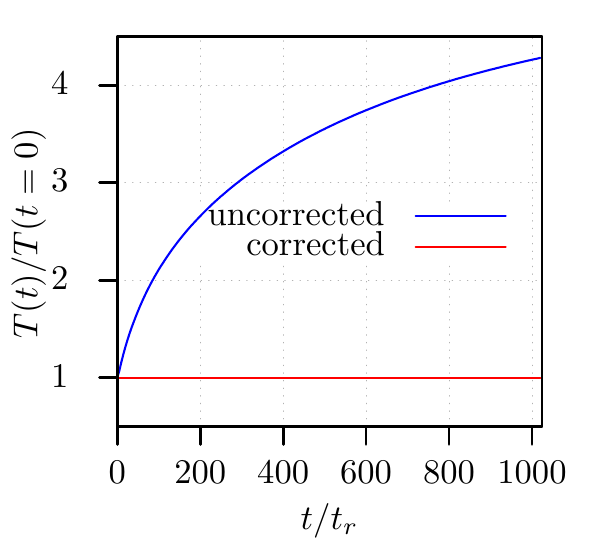}}
  \subcaptionbox{velocity}{\includegraphics[width=.32\textwidth]{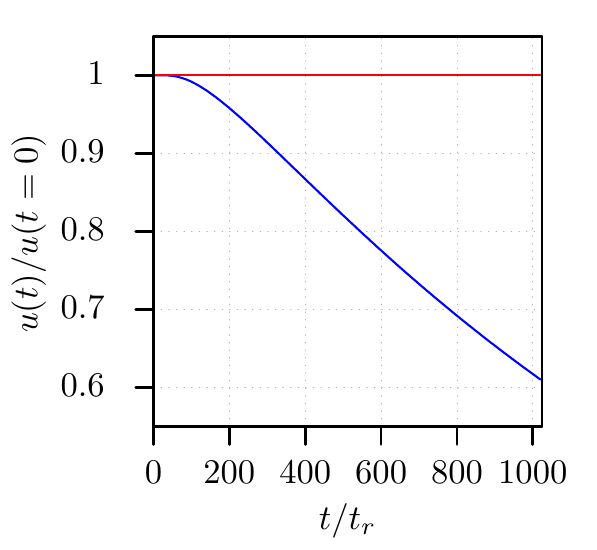}}
  \caption[Conservation in fundamental test (Maxwellian-like velocity
  distribution)]%
  { Growth of temperature $T$ and velocity $u$ over time $t$ in the
    fundamental test with a slightly decentered Maxwellian-like distribution in
    velocity space. The normalized density (a), temperature (b) and absolute
    velocity (c) are depicted for the uncorrected and corrected case. Time is
    normalized to the time for one rotation, $t_r$. }
  \label{fig:fundamental-test-m-T-v}
\end{figure}
In figure \ref{fig:fundamental-test-m-T-v}, the scalar temperature
$T \sim \mathcal E - m n \vec u^2/2$ (cf.~\eqref{eq:temperature-tensor}) and absolute
velocity $u = |\vec u|$ (cf.~\eqref{eq:momentum-density}) are plotted against time
for the uncorrected and corrected case, where the correction is enforced every
step on $T$ and $u$. As expected, both observables stay on a constant level.
\begin{figure}
  \centering
  \includegraphics[width=.8\textwidth]{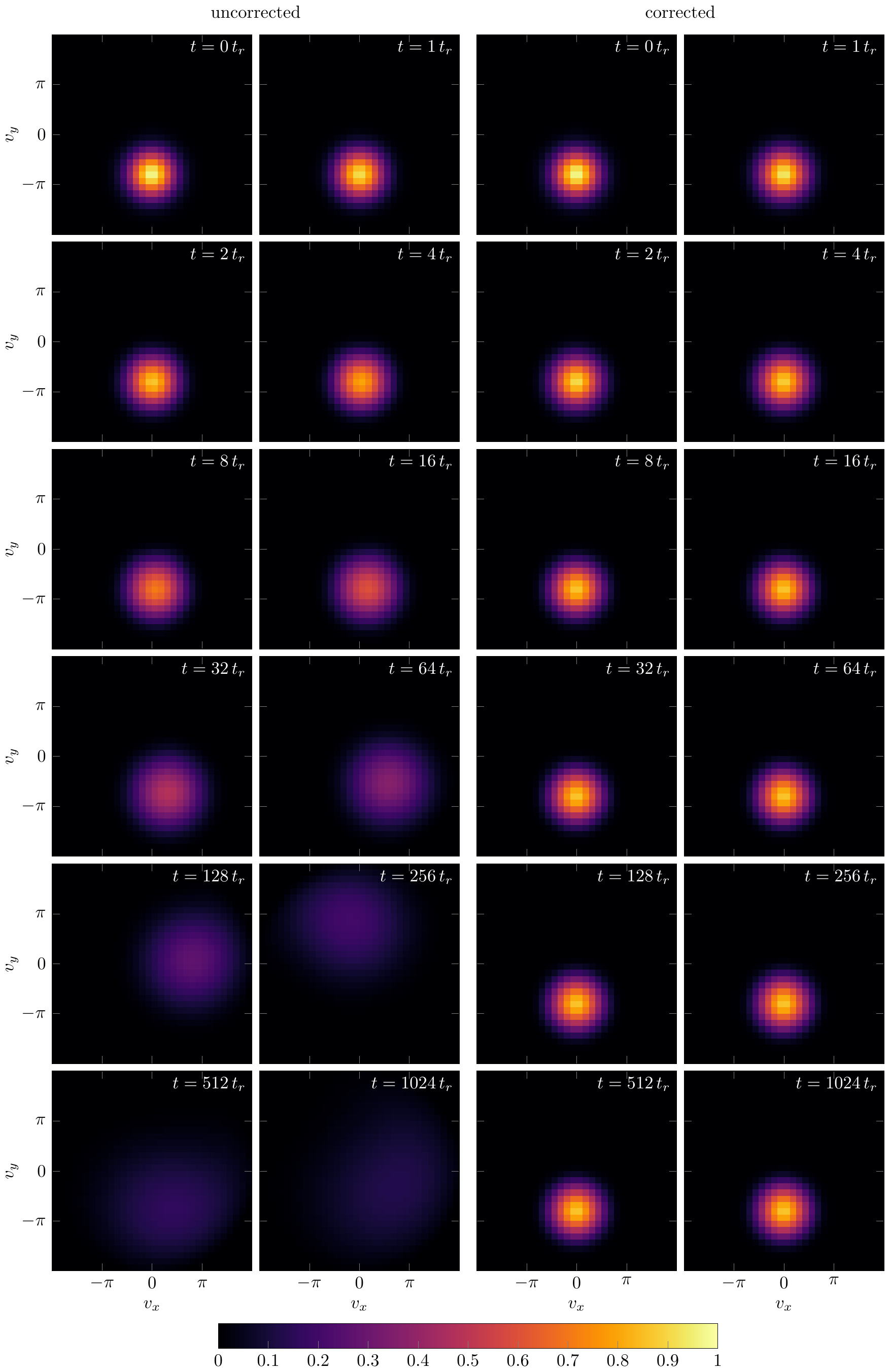}
  \caption[Phase-space density evolution in fundamental test Maxwellian-like
  velocity distribution)]%
  { Evolution of the phase-space density over time in the fundamental test with
    a slightly decentered Maxwellian-like distribution in velocity space. The
    correction is done stepwise on velocity and temperature. Time is normalized
    to the time for one rotation, $t_r$. }
  \label{fig:fundamental-test-m-psd}
\end{figure}
In figure \ref{fig:fundamental-test-m-psd}, the progression of the distribution
is depicted. Its initial peak value is normalized to one. The uncorrected distribution
apparently diffuses with time, but it stays constant in the corrected case.

As a second test case, the evolution of a completely non-Maxwellian,
pacman-shaped, rotating, but slightly decentered velocity distribution is
observed in two-dimensional $(v_x,v_y)$-space.
\begin{figure}
  \centering
  \subcaptionbox{density}{\includegraphics[width=.32\textwidth]{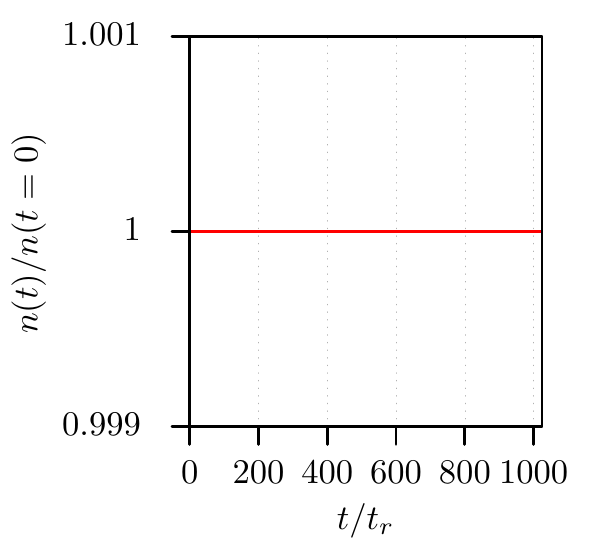}}
  \subcaptionbox{temperature}{\includegraphics[width=.32\textwidth]{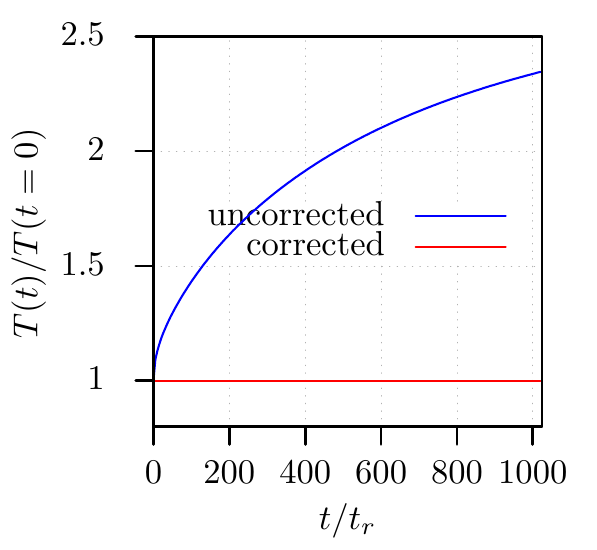}}
  \subcaptionbox{velocity}{\includegraphics[width=.32\textwidth]{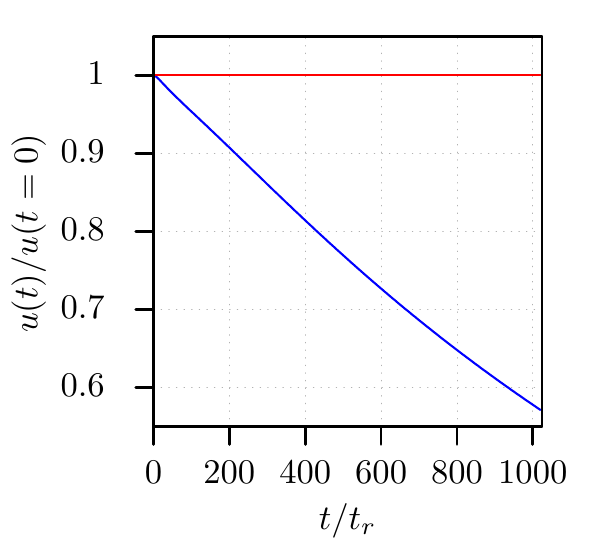}}
  \caption[Conservation in fundamental test (pacman-like velocity
  distribution)]%
  { Growth of temperature $T$ and velocity $u$ over time $t$ in the
    fundamental test with a slightly decentered pacman-like distribution in
    velocity space. The normalized density (a), temperature (b) and absolute
    velocity (c) are depicted for the uncorrected and corrected case. Time is
    normalized to the time for one rotation, $t_r$. }
  \label{fig:fundamental-test-p-T-v}
\end{figure}
In figure \ref{fig:fundamental-test-p-T-v}, the scalar temperature $T$ and
absolute velocity $u$ are again plotted against time for the uncorrected and
corrected case. Still, both observables stay on a constant level.
\begin{figure}
  \centering
  \includegraphics[width=.8\textwidth]{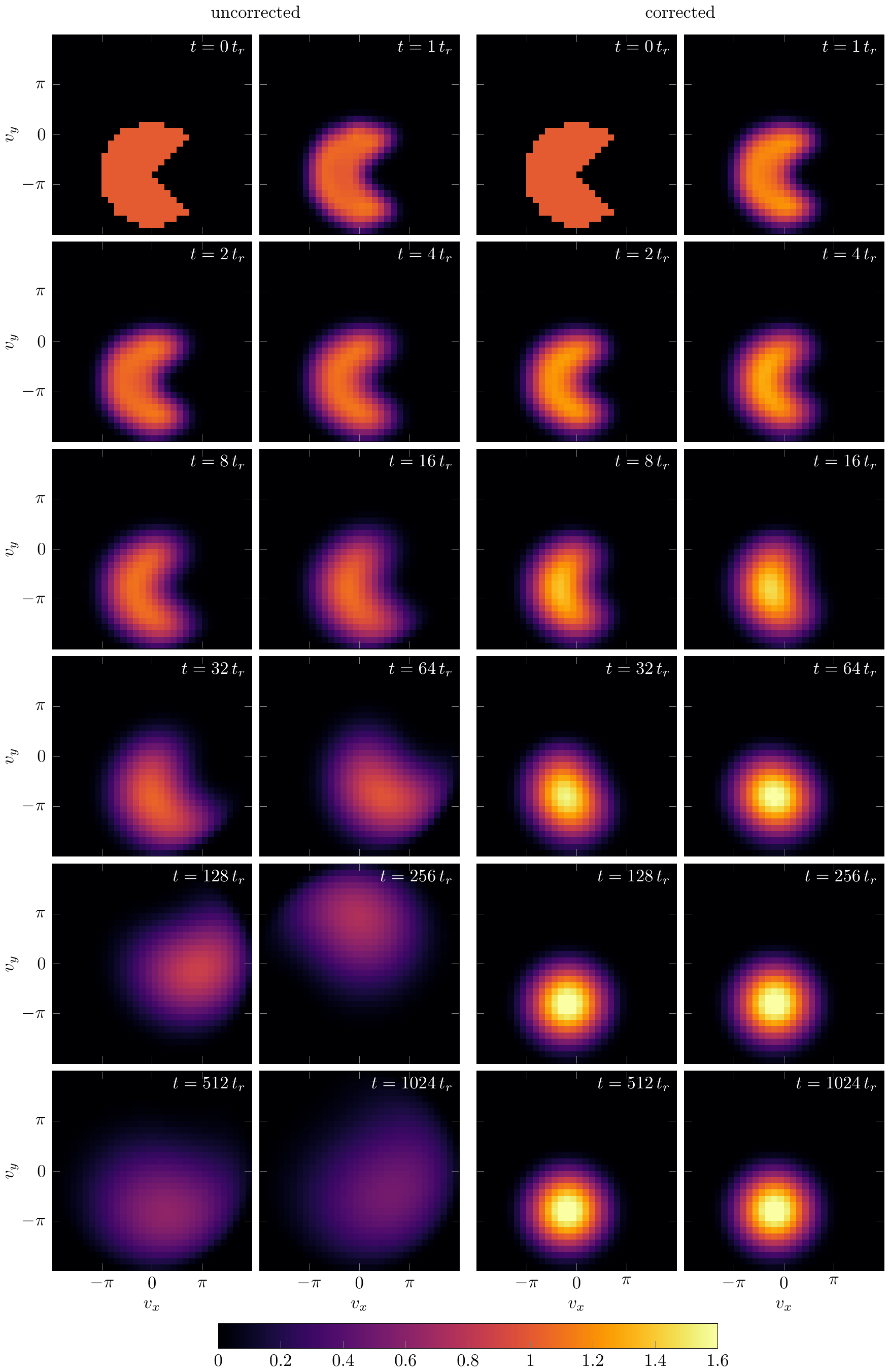}
  \caption[Phase-space density evolution in fundamental test (pacman-like
  velocity distribution)]%
  { Evolution of the phase-space density over time in the fundamental test with
    a slightly decentered pacman-like distribution in velocity space. The
    correction is done stepwise on velocity and temperature. Time is normalized
    to the time for one rotation, $t_r$. }
  \label{fig:fundamental-test-p-psd}
\end{figure}
In figure \ref{fig:fundamental-test-p-psd}, the progression of the distribution
is depicted with its initial value normalized to one. The rotation is clockwise.
Thus, in the uncorrected case, the rotation is a bit slower than expected from
the initial values. The pacman diffuses with time and becomes Maxwellian-like,
but its density basically stays the same in the corrected case.

\subsection{Fast magnetosonic wave}
The fast magnetosonic wave is a compressible wave with anisotropic
pressure that propagates perpendicular to the magnetic field and
occurs in the MHD regime. It has a phase velocity of
\begin{equation}
  c_\mathrm{fm}^2 = v_\mathrm{A}^2 + \frac{3k_\mathrm{B}(T_{0,e}+T_{0,i})}{m_i}
\end{equation}
where $v_\mathrm{A}$ is the Alfv\'en velocity. The initial conditions are
\begin{subequations}\label{eqs:fast-magnetosonic-init}
  \begin{align}
    n_s &= n_0\left(1 + \frac{\delta v}{c_\mathrm{fm}} \cos(\omega t - kx)\right) \\
    \vec{u}_s &= \delta v \cos(\omega t - kx)\,\hat{\vec x} \\
    \vec J &= \vec 0 \\
    \vec B &= B_0\left(1 + \frac{\delta v}{c_\mathrm{fm}} \cos(\omega t - kx)\right)\hat{\vec y} \\
    \vec E &= -B_0\delta v\cos(\omega t - kx)\,\hat{\vec z} \\
    p_{xx,s} &= p_{0,s} \left(1 + 3\frac{\delta v}{c_\mathrm{fm}} \cos(\omega t - kx)\right) \\
    p_{yy,s} = p_{zz,s} &= p_{0,s} \left(1 + \frac{\delta v}{c_\mathrm{fm}} \cos(\omega t - kx)\right)
  \end{align}
\end{subequations}
The correction of the phase space density by means of ten-moment
fitting is discussed on the basis of three simulations with similar
parameters. One simulation is done with the pure Vlasov model, one
with a pure ten-moment model and the last one with the Vlasov model in
combination with the ten-moment model for correction. All parameters
are consistent over the different simulations; the only differences
are due to the fact that some of the parameters do not have a meaning
in each of the configurations.

The simulation spans one transit of the wave through a periodic box
with a size of 22.4 ion inertial lengths times 11.2 ion inertial
lengths. In order to test the full two-dimensional code with this
essentially one-dimensional wave, the direction of propagation is not
parallel to the coordinate axes but skewed by an angle of
$\operatorname{tan}^{-1}(2)$. With a wave length of ten ion inertial
lengths and a phase velocity of twice the Alfv\'en velocity, the
overall setup is thus still periodic, but in a non-trivial way. The
speed of light is chosen to be $30\,v_\mathrm{A}$, the mass ratio is
$m_i/m_e = 25$, and the ion temperature is equal to the electron
temperature. In the pure ten moment run, the value for $k_0$ is five
inverse ion inertial lengths. The velocity spaces of ions and
electrons are cubes with and edge length of ten ion inertial lengths
and 20 ion inertial lengths, respectively. The spatial resolution is
$64\times 32$ cells, while there are $32^3$ cells in each velocity
space. The parameter $\varepsilon$ from the CWENO scheme is set to 0.1.
\begin{figure}
  \subcaptionbox{total energy}{\includegraphics{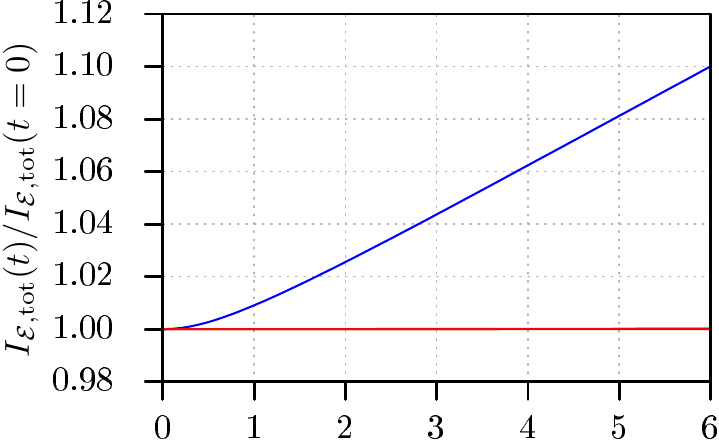}}
  \subcaptionbox{total energy (details)}{\includegraphics{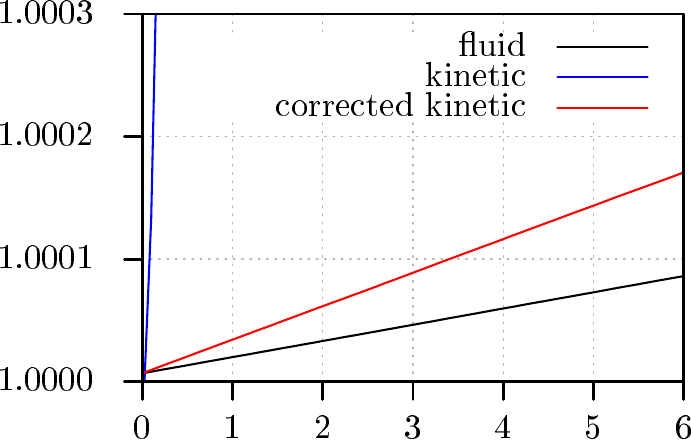}}

  \subcaptionbox{partition among species\label{fig:fast-magnetosonic-e-th}}{\hfill\includegraphics{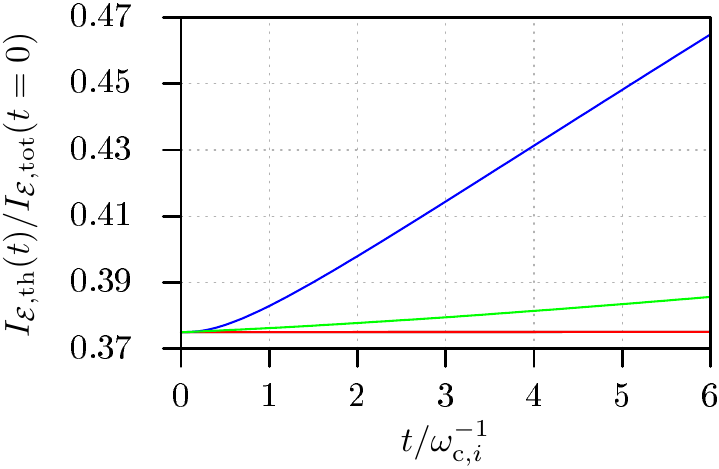}}
  \subcaptionbox{partition (details)\label{fig:fast-magnetosonic-e-th-details}}{\hfill\includegraphics{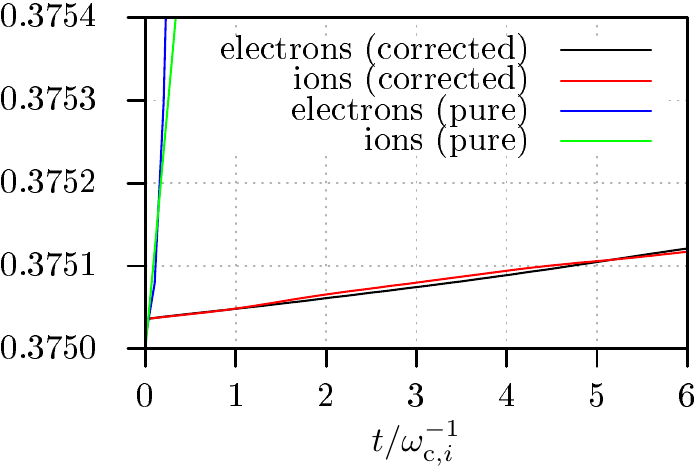}}
  \caption[Energy conservation in fast magnetosonic wave problem]%
  { Growth of energy over time in simulations of the fast magnetosonic
    wave with different models. The normalized total energy of the
    overall configuration is depicted in (a) and (b) for the three
    runs discussed in this section. In (c) and (d) the percental
    partition of thermal energy among electrons and ions is plotted
    for the pure kinetic run and the corrected kinetic run.}
  \label{fig:fast-magnetosonic-energy}
\end{figure}

\begin{figure}
  \subcaptionbox{$\tensor{E}_{xx,e}/\isunit{\mathcal E}$ (kinetic)}{\includegraphics{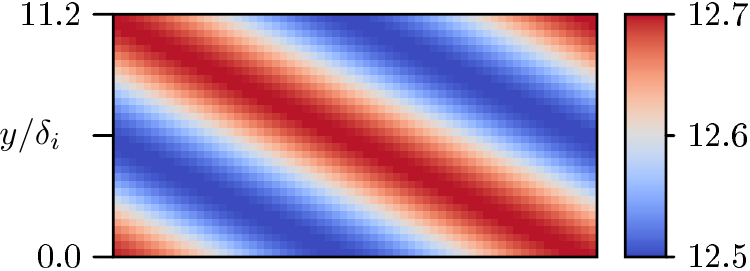}}
  \subcaptionbox{$\tensor{E}_{xy,e}/\isunit{\mathcal E}$ (kinetic)}{\hspace{2mm}\includegraphics{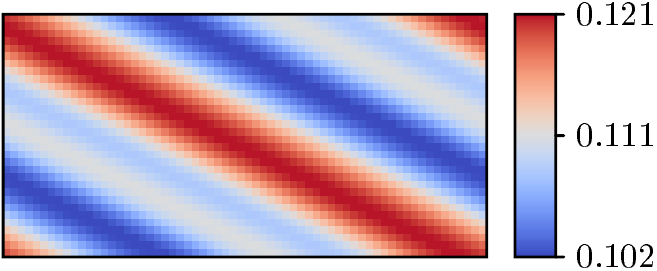}\hphantom{-}}

  \subcaptionbox{$\tensor{E}_{xx,e}/\isunit{\mathcal E}$ (kinetic, corrected)}{\includegraphics{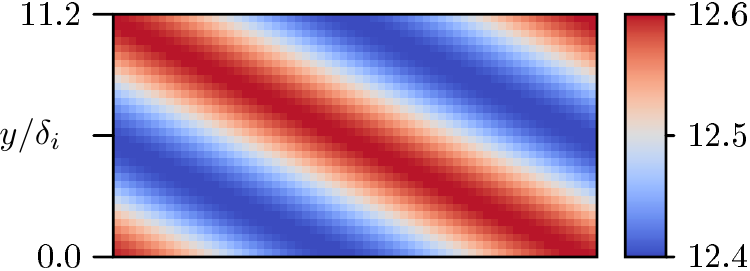}}
  \subcaptionbox{$\tensor{E}_{xy,e}/\isunit{\mathcal E}$ (kinetic, corrected)}{\hspace{2mm}\includegraphics{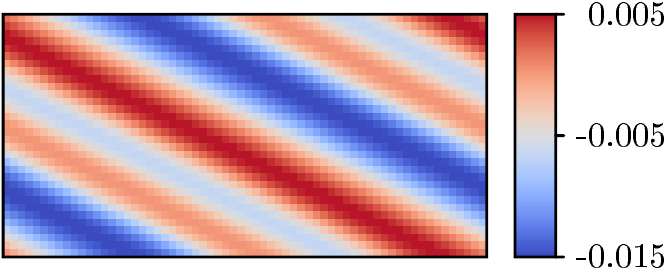}}

  \subcaptionbox{$\tensor{E}_{xx,e}/\isunit{\mathcal E}$ (fluid)}{\includegraphics{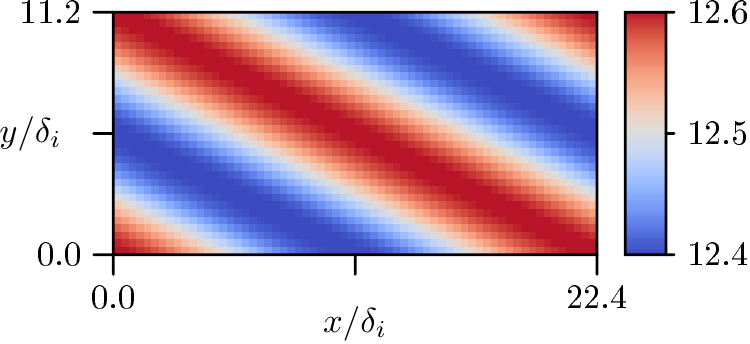}}
  \subcaptionbox{$\tensor{E}_{xy,e}/\isunit{\mathcal E}$ (fluid)}{\includegraphics{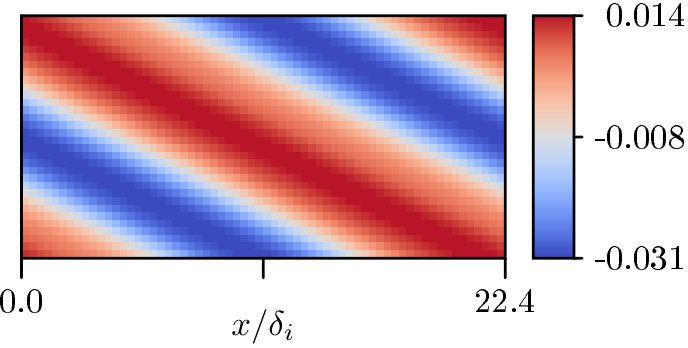}}
  \caption[Energy in fast magnetosonic wave]%
  { Selected components of the electron energy tensor $\tensor{E}_e$
    after one wave-transit in simulations of the fast magnetosonic
    wave with different models.}
  \label{fig:fast-magnetosonic-e}
\end{figure}

As it can be seen in \eqref{fig:fast-magnetosonic-energy}, the
numerical heating actually is an issue in the pure kinetic
simulation. The integral of the total energy $I_{\mathcal E,
  \mathrm{tot}}$ rises by about 10\% over the course of the
simulation. In \ref{fig:fast-magnetosonic-e-th}, where the
analogously defined thermal part $I_{\mathcal E, \mathrm{th}}$ of
$I_{\mathcal E, \mathrm{tot}}$ and its distribution among the species
in the plasma is plotted, it becomes apparent that by far the largest
part of the energy growth is due to an increase in the inner energy of
the electrons. This and the fact that the rise is almost linear is an
indication that the heating is a consequence of the gyro motion and
the concomitant smearing of the distribution function, just as it
could be expected. The lighter electrons gyrate faster and are thus
more susceptible for this effect.

The correction with ten-moment fitting leads to energy conservation
that is almost as good as that of the pure fluid simulation. It must
be kept in mind that for reasons of efficiency on the graphics cards
only single precision floating point numbers are used, so that a
better result can hardly be expected. Interestingly, there is
practically no difference with respect to energy conservation between
electrons and ions, when fitting is used
(\ref{fig:fast-magnetosonic-e-th-details}). The mass ration seems to
play no role in that case.

A view on the actual results of the simulations
(\ref{fig:fast-magnetosonic-e}) reveals noticeable differences
between the simulations in the more critical quantities like the
off-diagonal components of the energy tensor. The corrected kinetic
run appears to lie in between the pure kinetic and the pure fluid run,
but from the few low-resolution runs that were made it is hard to
judge which of the runs is closer to the right result. In any way, it
cannot be expected that the kinetic model leads to the same results as
the ten-moment model that is artificially closed at the level of the
heat flux and relies on additional parameters. Besides that, a small
amplitude wave is probably not the best candidate for identifying
actual problems with the behavior of the different schemes. The other
results below are better suited for this kind of analysis.

In summary, the correction with ten-moment fitting leads to remarkably
good conservation properties and at least seems to not introduce
obviously wrong side-effects.

\subsection{GEM challenge}
The GEM challenge \cite{bir2001} is a well-studied problem for
examining magnetic reconnection. The initial setup is a Harris sheet
\cite{har1962} that is periodic in the $x$-direction:
\begin{subequations}\label{eqs:gem-init-cond}
  \begin{align}
    \vec E &= \vec 0 \\
    \vec B &= B_0 \tanh\left(\frac{y}{\lambda}\right)\hat{\vec x} \\
    n_s &= n_\mathrm{bg} + n_0 \operatorname{sech}^2\left(\frac{y}{\lambda}\right) \\
    \vec{u}_s &= -\frac{\Theta_sB_0}{q_s \mu_0 \lambda n_0}\left(\frac{1}{1+\frac{n_\mathrm{bg}}{n_0}\cosh^2\left(\frac{y}{\lambda}\right)}\right)\hat{\vec z} = \frac{\delta_i} \lambda \left(\frac{-\operatorname{sgn}(q_s)\Theta_s}{1+\frac{n_\mathrm{bg}}{n_0}\cosh^2\left(\frac{y}{\lambda}\right)}\right)v_\mathrm{A}\hat{\vec z}\\
    k_\mathrm{B}T_{yy,s} &= \frac{\Theta_sB_0^2}{2\mu_0n_0} = \frac 1 2 \Theta_s m_i v_\mathrm{A}^2
  \end{align}
\end{subequations}
This sheet is edged with conducting walls above and
below. Reconnection is triggered with a magnetic island perturbation
of the form:
\begin{equation}
  \delta \vec B = \hat{\vec z}\times\nabla\left(\delta \psi \cos\left(\frac{2\pi x}{L_x}\right)\cos\left(\frac{\pi y}{L_y}\right)\right) = \delta \psi \begin{pmatrix}-\frac{\pi}{L_y}\cos\left(\frac{2\pi x}{L_x}\right)\sin\left(\frac{\pi y}{L_y}\right) \\ \phantom{-}\frac{2\pi}{L_x}\sin\left(\frac{2\pi x}{L_x}\right)\cos\left(\frac{\pi y}{L_y}\right) \\ 0\end{pmatrix} \label{eq:perturbation-gem}
\end{equation}
A variety of simulations of the GEM challenge are done. While the
physical and most of the numerical parameters are kept constant as far
as possible, the models, some methods, and certain numerical
parameters are varied. The runs with the distinguishing parameters and
names for later reference are listed in tables
(\ref{tab:gem-parameters_a}, \ref{tab:gem-parameters_b}). They differ with respect to the combination
of physical models, the way in which the fitting matrix $\mat A$ is
calculated, and the parameter $\varepsilon$, which determines the
amount of numerical diffusivity in the CWENO scheme referenced above
in \ref{sec:fluid-solver}. A larger $\varepsilon$ corresponds to less
artificial diffusivity.

In all simulations, the physical parameters are the same: The domain
has a size of $8\pi\times 4\pi$ ion intertial lengths, the ratio
between the speed of light and the Alfv\'en velocity is 20 and the
amplitude of the initial magnetic perturbation is
$0.1\,B_0$. Furthermore, $m_i/m_e = 25$ and $T_i/T_e = 5$. In the
Vlasov simulations, the velocity spaces are cubes with an edge length
of 24 ion inertial lengths for electrons and 10 ion inertial lenghts
for ions. In the pure ten-moment runs, the parameter $k_0$ introduced
in \eqref{eq:closure-ten-moment} is set to five inverse ion inertial
lengths, following \citet{wang-et-al:2015}. All runs are done with a
resolution of $256\times 128$ cells in physical space. If relevant,
the velocity space is resolved with $32^3$ cells at each point in
space.
\begin{table}[h]
  \begin{tabularx}{1.0\textwidth}{XXP{1.4cm}P{2.3cm}}
    \toprule
    name & model & $\varepsilon$ & calculation of $\mat A$ \\
    \midrule
    GEM-5-S    & pure five-moment fluid &  $\num{1e-6}$ & -- \\
    GEM-5-M    & pure five-moment fluid &  0.1 & -- \\
    GEM-10-S   & pure ten-moment fluid &  $\num{1e-6}$ & -- \\
    GEM-10-M   & pure ten-moment fluid  &  0.1 & -- \\
    GEM-V      & pure Vlasov model &  -- & --  \\
    GEM-V5-S   & Vlasov + five-moment fluid &  $\num{1e-6}$ & -- \\
    GEM-V5-M   & Vlasov + five-moment fluid &  0.1 & -- \\
    GEM-V10-SG & Vlasov + ten-moment fluid &  $\num{1e-6}$ & gradient descent \\
    GEM-V10-MG & Vlasov + ten-moment fluid &  0.1 & gradient descent \\
    GEM-V10-SA & Vlasov + ten-moment fluid &  $\num{1e-6}$ & approximate \\
    GEM-V10-MA & Vlasov + ten-moment fluid &  0.1 & approximate \\
    \bottomrule
  \end{tabularx}
  \caption[Parameters for GEM challenge]%
  { Parameters for the simulations of the GEM challenge. The omissions
    indicate that the respective category has no counterpart in the
    respective setup; for example, there is no fitting matrix in a pure fluid run. }
  \label{tab:gem-parameters_a}
\end{table}

The different simulations of the GEM challenge can be compared in a
huge variety of ways. Here, some of these that are particularly
illustrative in one way or another are picked out in order to give a
clear picture.

\begin{figure}
  \includegraphics{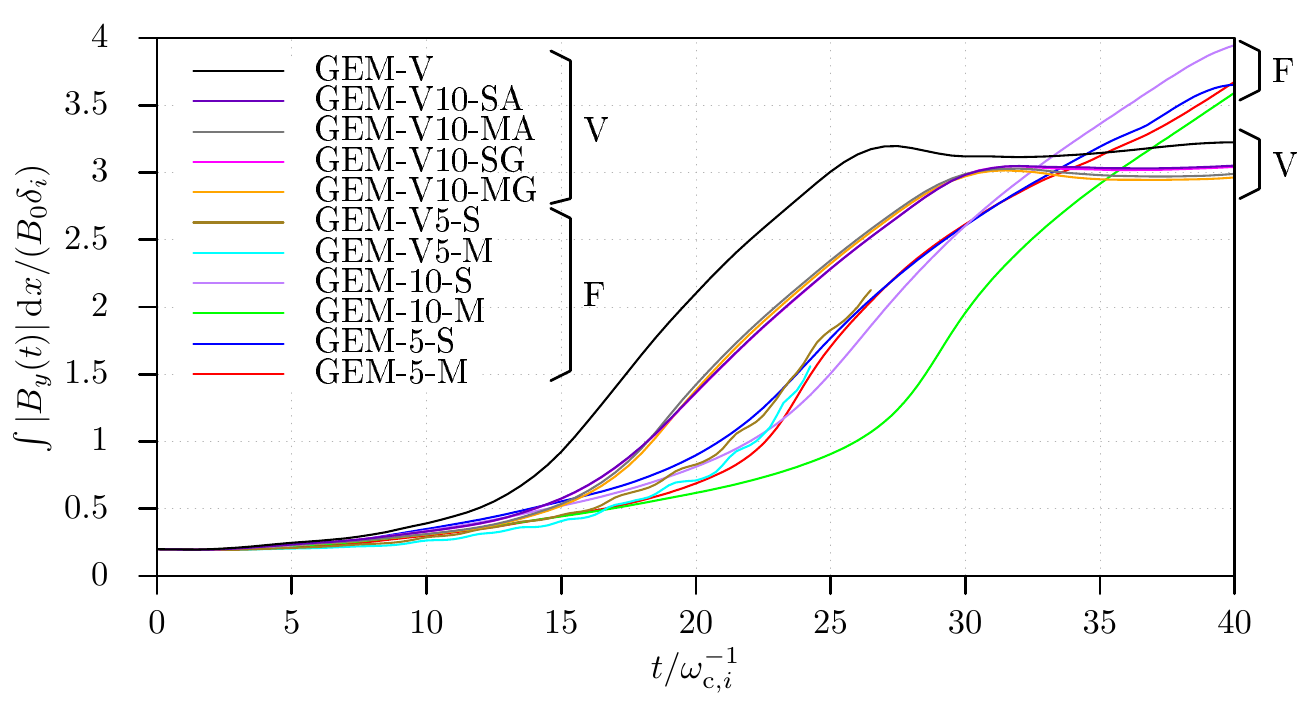}
  \caption[Reconnected flux in GEM challenge]%
  { Reconnected flux over time in the different simulations of the GEM
    challenge.}
  \label{fig:gem-challenge-reco-flux}
\end{figure}

As often in reconnection and for making the comparison with the data
from the literature possible, the first thing to look at is the
reconnected magnetic flux over time. For the current configuration,
this flux can be measured as the integral of $|B_y|$ along the current
sheet. Plots of this quantity over time are given in
\ref{fig:gem-challenge-reco-flux} for all simulations. While the
growth rates all indicate fast reconnection and lie around a value of
$\myunit{0.2}{B_0v_\mathrm{A}}$, which is comparable to the results
presented by \citet{bir2001}, the beginning of the process is
at very different times and (at least) two groups of runs can clearly
be distinguished: The flux in the simulations of the first group, the
pure kinetic simulation and the kinetic simulations with ten-moment
fitting, saturates at a value of around $\myunit{3}{B_0\delta_i}$ and
at a time between 25 and 30 inverse ion gyro periods, which is very
similar to what is shown for hybrid and particle runs in
\cite{bir2001}. The second group consists of all other
runs. The reconnected flux in these simulations does not saturate over
the given period of time and resembles the Hall MHD curve in
\cite{bir2001}, which might partly be due to the low
resolution that may suppress some two-fluid effects. The Vlasov runs
with five-moment fitting have different curves than the simulations of
the first groups and exhibit signs of artificial oscillations. They
could not even be completed due to onsetting numerical instabilities.

There are at least three important results from this first comparison:
Firstly, the ten-moment fitting correction delays the commencement of
reconnection and leads to saturation at a slightly lower level, but
the shape of the resulting curve is very similar to the kinetic one. With the pure
ten-moment model, the onset of reconnection is significantly
later. Secondly, all methods of ten-moment fitting that are tested
lead to basically the same curve, which is remarkable, when compared
to the variation between the two pure ten-moment runs that only differ
with respect to the numerical parameter $\varepsilon$. Such
differences seem to play a minor role when fitting is
employed. Thirdly, five-moment correction fitting might work for
simple test cases, but for general problems with non-linear dynamics
isotropic fitting appears to influence the simulation heavily and
render it invalid.

\begin{figure}
  \subcaptionbox{Overview.}{\includegraphics{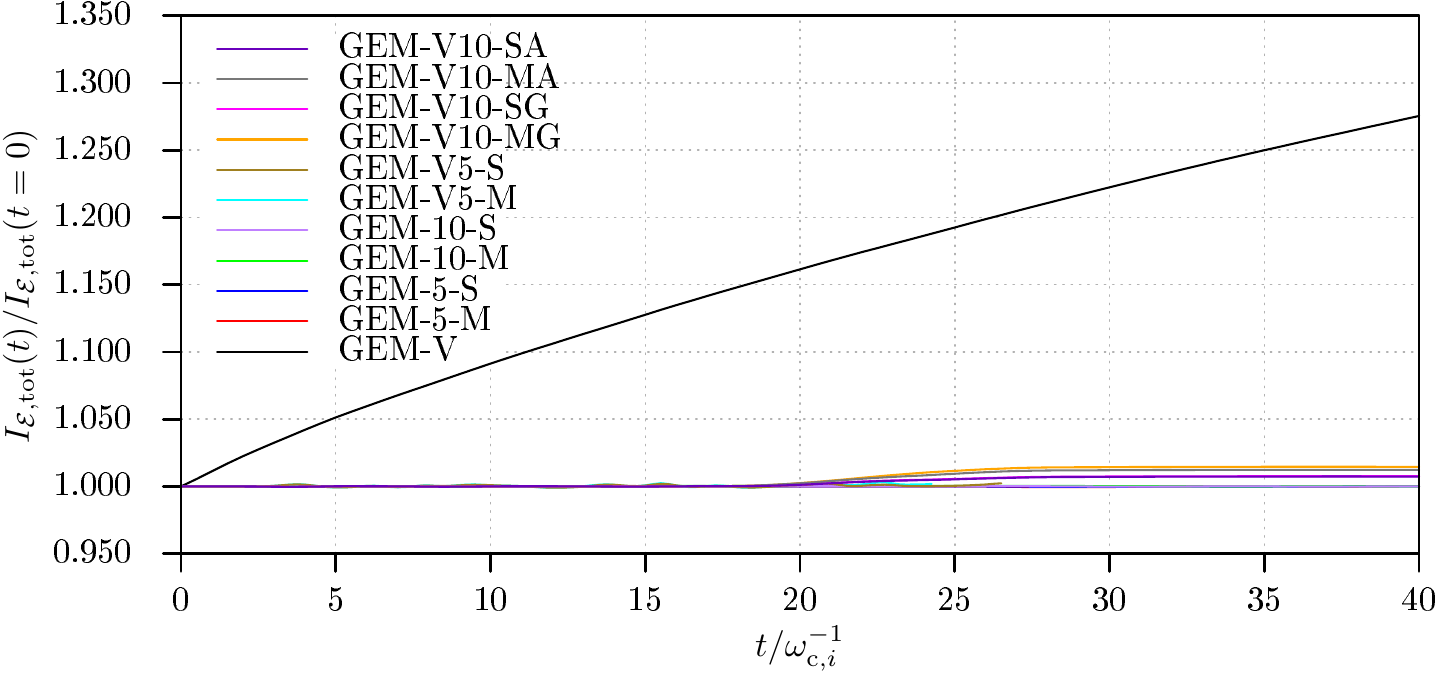}}
  \subcaptionbox{Details.}{\includegraphics{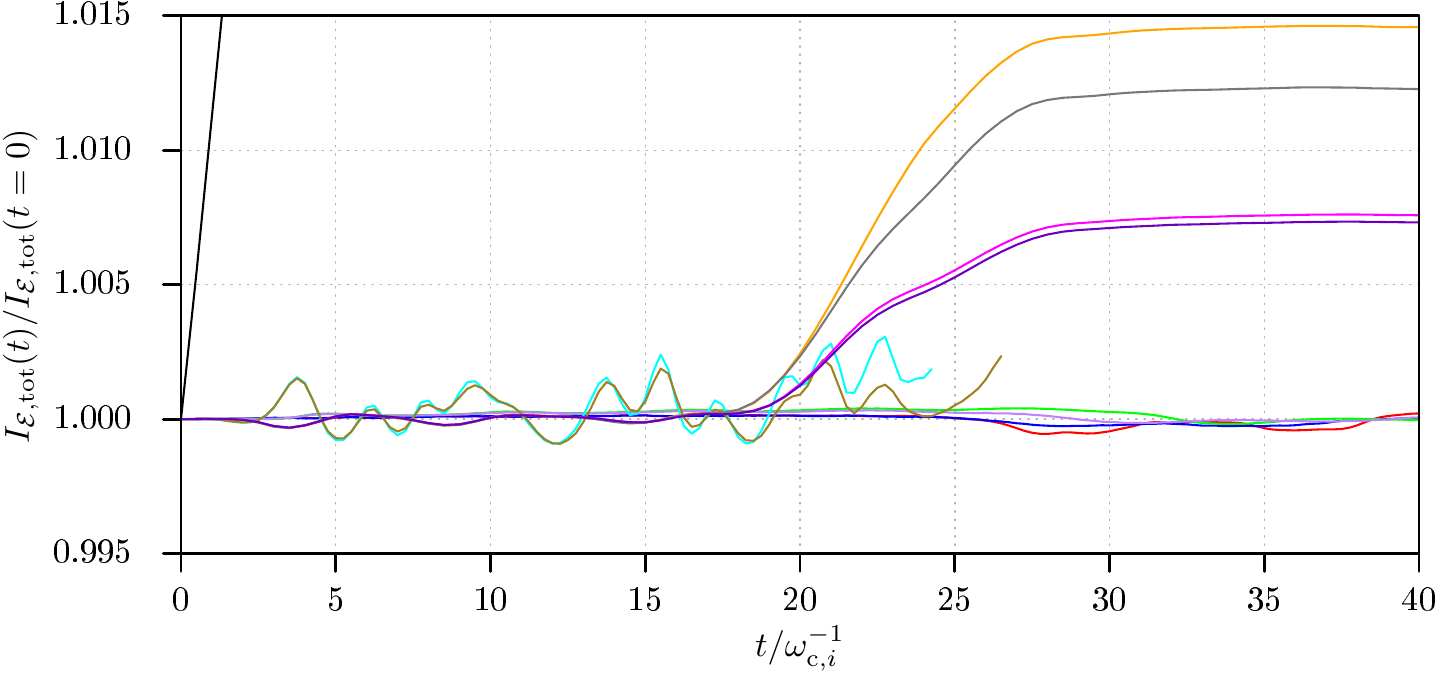}}
  \caption[Energy conservation in GEM challenge]%
  { Normalized total integrated energy over time in different
    simulations of the GEM challenge. For the exact definition of the
    total energy $I_{\mathcal E, \mathrm{tot}}$ including the kinetic particle
    energy and the contributions from the electromagnetic fields. Because of the large differences
    between the ranges of the curves, two plots of the same data with
    different scaling are given.}
  \label{fig:gem-challenge-energy}
\end{figure}

The main purpose of correction fitting is to get better conservation
properties and in particular less artificial heating in the Vlasov
code. This can be checked by examining the integral of the total
energy $\mathcal{E}_\mathrm{tot}$ over the whole domain that is
plotted depending on the time and for the different runs in
\ref{fig:gem-challenge-energy}. Here, the shortcomings of five-moment
fitting become even more apparent than in
\ref{fig:gem-challenge-reco-flux}. Non-physical oscillations are
present almost from the beginning and the curves from the
five-moment fitting runs do not resemble the results from the more
plausible simulations in any way.  Besides that, independently of the
way in which ten-moment fitting is done, energy conservation is
significantly improved through the correction routines. While the
overall energy rises by about 27\% in the pure kinetic simulation, the
overall increase in energy hardly exceeds 1\% for all other runs. This
is not as good as the almost perfect conservation in the fluid codes
or the results from the fast magnetosonic wave, but it must be
taken into account that by far the majority of the energy increase
stems from the phase around the time of the peak reconnection rate,
when the conditions are more extreme, because all quantities vary
faster and $f_s$ exhibits a higher anisotropy. Contrary to the poor
discriminability of the curves of the ten-moment fitting runs in
\ref{fig:gem-challenge-reco-flux}, both $\varepsilon$ and the method for
calculating $\mat A$ make a difference with respect to energy
conservation. A smaller $\varepsilon$ and an approximate calculation of
$\mat A$ appear to lead to better energy conservation. The last point
is not very unambiguous for the small $\varepsilon$, but it might be
explained with the more noisy values of $\mat A$ that happen to be the
result of the gradient descent method.

\begin{figure}
  \subcaptionbox{GEM-10-M ($t=\myunit{28.75}{\omega_{\mathrm{c},i}^{-1}}$)}{\includegraphics{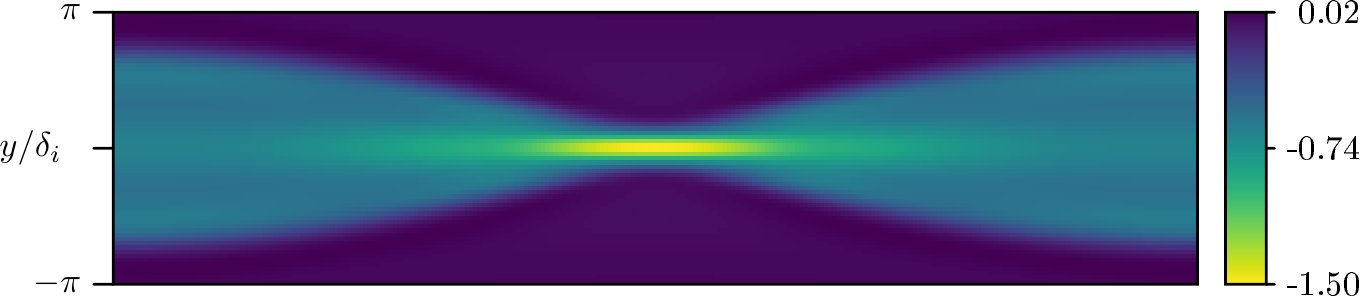}}
  \subcaptionbox{GEM-V10-MA ($t=\myunit{20.50}{\omega_{\mathrm{c},i}^{-1}}$)}{\includegraphics{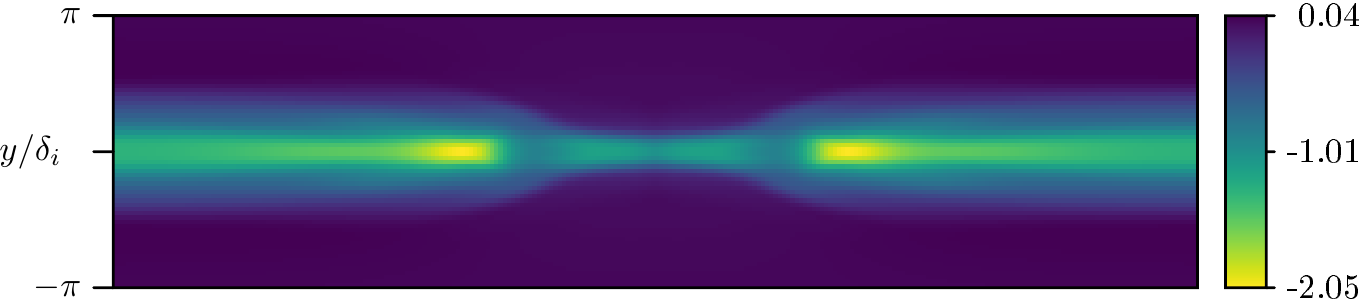}}
  \subcaptionbox{GEM-V ($t=\myunit{17.50}{\omega_{\mathrm{c},i}^{-1}}$)}{\includegraphics{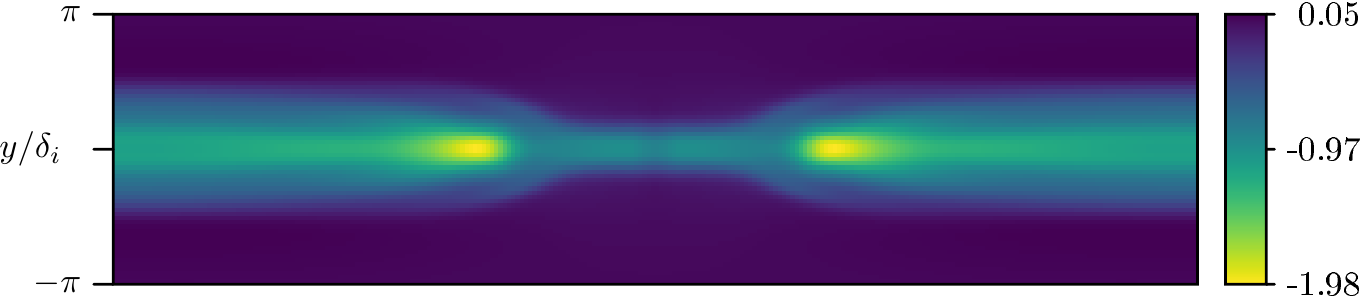}}
  \subcaptionbox{GEM-V5-M ($t=\myunit{24.25}{\omega_{\mathrm{c},i}^{-1}}$)}{\includegraphics{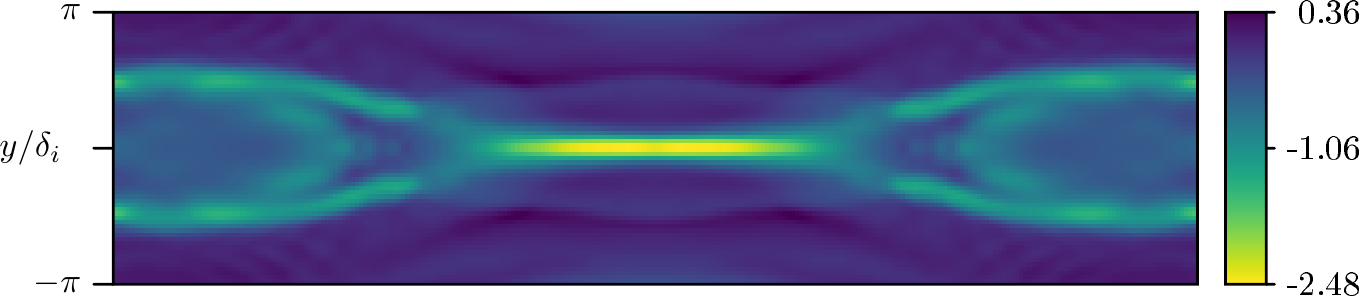}}
  \subcaptionbox{GEM-5-M ($t=\myunit{24.25}{\omega_{\mathrm{c},i}^{-1}}$)}{\includegraphics{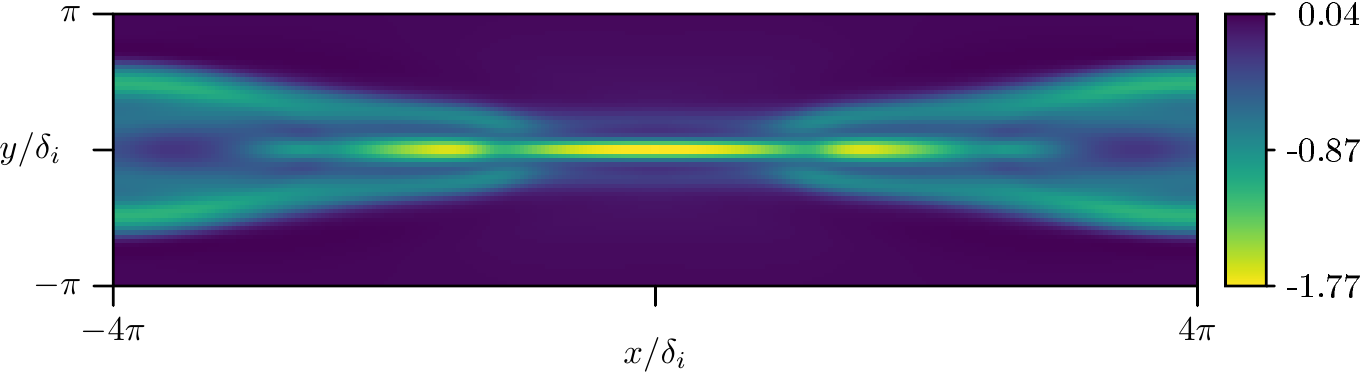}}
  \caption[Comparison of $J_z$ in GEM challenge]%
  { $J_z/(en_0v_\mathrm{A})$ in GEM challenge; comparison of different
    numerical schemes and models at time of comparable reconnected
    flux, roughly at time of highest reconnection rate (\cf
    \ref{fig:gem-challenge-reco-flux}).}
  \label{fig:gem-challenge-jz-comparison}
\end{figure}

The reconnection curves and the energy curves are good for an
overview, but they hide the details of the mechanisms with which they
were generated. In order to get a better picture of qualitative
differences between the runs, the structure of the current sheet, \ie
the value of $J_z$, is given for times of comparable reconnected flux
in \ref{fig:gem-challenge-jz-comparison}. Here, the differences
between the runs become more apparent: GEM-V10-MA exhibits a very
similar structure to that in GEM-V. In both of them, the current sheet
is split apart and $J_z$ is concentrated in two channels at
the end points of the respective gap. Nevertheless, the velocities in
$z$-direction and in particular $v_{z,e}$ are still peaked around the
X-point. The two pure fluid runs that are shown are very different;
they developed a clearer global X-point structure with a distinct
sheet at the reconnection site. The most interesting observation is
that GEM-V5-M is closer to the fluid runs than to the pure kinetic
ones, whereby it does not fall clearly into one of the two categories
and exhibits features that cannot be found in the other plots.

\begin{figure}
  \subcaptionbox{GEM-V ($t=18.50\omega_{\mathrm{c},i}^{-1}$)}{\includegraphics{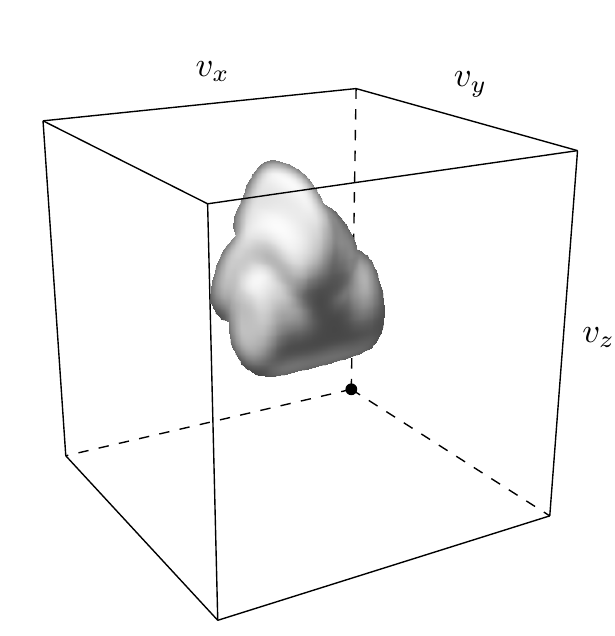}}
  \subcaptionbox{GEM-V10-MG ($t=20.75\omega_{\mathrm{c},i}^{-1}$)}{\includegraphics{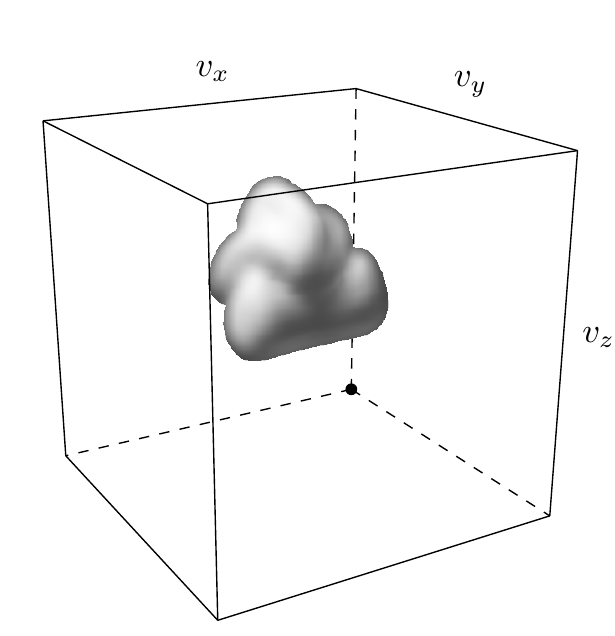}}
  \caption[Isosurface of electron phase space density in GEM challenge]%
  { Isosurface of $f_e(\vec v)$ in different simulations of GEM
    challenge during time of peak reconnection rate in vicinity of
    X-point at $\vec x = (x,y) \approx (0.15\delta_i, -0.15\delta_i)$.}
  \label{fig:gem-challenge-f3}
\end{figure}

So, while five-moment fitting introduces all kinds of side-effects,
ten-moment fitting appears to do what it is expected to: it improves
energy conservation while retaining the kinetic mechanisms. For a
further check of this, consider the electron phase space density near
the X-point at the time of peak reconnection rate
\ref{fig:gem-challenge-f3}. There is a clear qualitative similarity
between the shapes of the isosurfaces, but they are far from being
equal. The two surfaces might be described as distorted versions of
each other. At this level it is hard to say which of the two is more
correct; other metrics like the conservation properties discussed
above are probably better for judging the overall quality of the
schemes. Anyway the results from the fitting run do not appear to be
implausible in comparison to that of the pure kinetic run and on a
qualitative level they are basically the same.

In summary it can be said that ten-moment fitting significantly
improves the conservation properties of the Vlasov code without
introducing spurious extra-effects. Approximate calculation of the
fitting matrix $\mat A$ seems to be the better idea and a smaller
$\varepsilon$ in the CWENO scheme leads to a lower increase in total
energy, but in particular the last point is not very strict and the
actual differences between the tested methods of ten-moment fitting
are remarkably small, both in the data shown and in the other
quantities that were not discussed in detail. On the other hand,
five-moment fitting does not work well. Forcing the distribution
function to match the moments of an essentially isotropic fluid model
apparently destroys essential kinetic features, leads to artificial
instabilities and renders the overall simulation useless.

\begin{table}[h]
  \begin{tabularx}{1.0\textwidth}{XP{1.9cm}P{1.2em}P{3.4em}P{3.4em}P{2.3em}P{2.em}P{2.8cm}}
    \toprule
    name & resolution & $n_v$ & $L_{v,e}/v_\mathrm{A}$ & $L_{v,i}/v_\mathrm{A}$ & $\varepsilon$ & $\delta_ik_0$ & calculation of $\mat A$ \\
    \midrule
    GEM-5-S & $256\times 128$ & -- & -- & -- & $\num{1e-6}$ & -- & -- \\
    GEM-5-M & $256\times 128$ & -- & -- & -- & 0.1 & -- & -- \\
    GEM-10-S & $256\times 128$ & -- & -- & -- & $\num{1e-6}$ & 5 & -- \\
    GEM-10-M & $256\times 128$ & -- & -- & -- & 0.1 & 5 & -- \\
    GEM-V & $256\times 128$ &  32 & 24 & 10 & -- & -- & -- \\
    GEM-V5-S & $256\times 128$ & 32 & 24 & 10 & $\num{1e-6}$ & -- & -- \\
    GEM-V5-M & $256\times 128$ & 32 & 24 & 10 & 0.1 & -- & -- \\
    GEM-V10-SG & $256\times 128$ & 32 & 24 & 10 & $\num{1e-6}$ & -- & gradient descent \\
    GEM-V10-MG & $256\times 128$ & 32 & 24 & 10 & 0.1 & -- & gradient descent \\
    GEM-V10-SA & $256\times 128$ & 32 & 24 & 10 &  $\num{1e-6}$ & -- & approximate \\
    GEM-V10-MA & $256\times 128$ & 32 & 24 & 10 &  0.1 & -- & approximate \\
    \bottomrule
  \end{tabularx}
  \caption[Parameters for GEM challenge]%
  { Parameters for the simulations of the GEM challenge. The omissions
    indicate that the respective category has no counterpart in the
    respective setup; for example, there is nothing like a resolution
    and size of a velocity in pure fluid simulations. }
  \label{tab:gem-parameters_b}
\end{table}

\section{Results and conclusion}\label{sec:conclusion}
A general method for enhancing the conservation properties of Vlasov
solvers working on a spatial grid is proposed that is not bound to a
specific numerical scheme. The method is tested on different test
problems that stress several aspects like the behavior in the linear
or the fully non-linear regime. It is found that the fully anisotropic
correction scheme gives satisfactory results and very good
conservation properties in all cases, while the related isotropic
fitting procedure, of which the first is a generalization, leads to
numerical instabilities and can thus not be used.

The next step is the application of the ten-moment fitting method to
the problem of coupling fluid and kinetic codes that live in separate
regions, as a generalization of the method presented in
\cite{rieke:2015}, also in combination with the correction method
shown here. The respective results will be published in a future paper.
\section*{Acknowledgment}
We acknowledge all the fruitful discussions we had with J\"urgen
Dreher.  This research was supported by the DFG Research Unit FOR
1048, project B2. Calculations were partly performed on the CUDA-Cluster
DaVinci hosted by the Research Department \textit{Plasmas with Complex
  Interactions} at the Ruhr-University Bochum. We gratefully
acknowledge the computing time granted by the John von Neumann
Institute for Computing (NIC) and provided on the supercomputer JURECA
at J\"{u}lich Supercomputing Centre (JSC).

\bibliography{lit}

\begin{thebibliography}{28}
\expandafter\ifx\csname natexlab\endcsname\relax\def\natexlab#1{#1}\fi
\providecommand{\url}[1]{\texttt{#1}}
\providecommand{\href}[2]{#2}
\providecommand{\path}[1]{#1}
\providecommand{\DOIprefix}{doi:}
\providecommand{\ArXivprefix}{arXiv:}
\providecommand{\URLprefix}{URL: }
\providecommand{\Pubmedprefix}{pmid:}
\providecommand{\doi}[1]{\href{http://dx.doi.org/#1}{\path{#1}}}
\providecommand{\Pubmed}[1]{\href{pmid:#1}{\path{#1}}}
\providecommand{\bibinfo}[2]{#2}
\ifx\xfnm\relax \def\xfnm[#1]{\unskip,\space#1}\fi
\bibitem[{Rieke et~al.(2015)Rieke, Trost, and Grauer}]{rieke:2015}
\bibinfo{author}{M.~Rieke}, \bibinfo{author}{T.~Trost},
  \bibinfo{author}{R.~Grauer},
\newblock \bibinfo{title}{Coupled {V}lasov and two-fluid codes on {GPU}s},
\newblock \bibinfo{journal}{Journal of Computational Physics}
  \bibinfo{volume}{283} (\bibinfo{year}{2015}) \bibinfo{pages}{436--452}.
\bibitem[{Arnold and Giering(1997)}]{arnold-giering:1997}
\bibinfo{author}{A.~Arnold}, \bibinfo{author}{U.~Giering},
\newblock \bibinfo{title}{{An analysis of the Marshak conditions for matching
  Boltzmann and Euler equations}},
\newblock \bibinfo{journal}{Math. Models Methods Appl. Sci.}
  \bibinfo{volume}{7} (\bibinfo{year}{1997}) \bibinfo{pages}{557--577}.
\bibitem[{Degond et~al.(2010)Degond, Dimarce, and
  Mieussens}]{degond-et-al:2010}
\bibinfo{author}{P.~Degond}, \bibinfo{author}{G.~Dimarce},
  \bibinfo{author}{L.~Mieussens},
\newblock \bibinfo{title}{{A multiscale kinetic-fluid solver with dynamic
  localization of kinetic effects}},
\newblock \bibinfo{journal}{J. Comput. Phys.} \bibinfo{volume}{229}
  (\bibinfo{year}{2010}) \bibinfo{pages}{4907--4933}.
\bibitem[{Dellacherie(2003)}]{dellacherie:2003}
\bibinfo{author}{S.~Dellacherie},
\newblock \bibinfo{title}{{Kinetic-Fluid Coupling in the Field of the Atomic
  Vapor Isotopic Separation: Numerical Results in the Case of a Monospecies
  Perfect Gas}},
\newblock \bibinfo{journal}{AIP Conf. Proc.} \bibinfo{volume}{663}
  (\bibinfo{year}{2003}) \bibinfo{pages}{947--956}.
\bibitem[{Goudon et~al.(2013)Goudon, Jin, Liu, and Yan}]{goudon-et-al:2013}
\bibinfo{author}{T.~Goudon}, \bibinfo{author}{S.~Jin}, \bibinfo{author}{J.-G.
  Liu}, \bibinfo{author}{B.~Yan},
\newblock \bibinfo{title}{{Asymptotic-preserving schemes for kinetic-fluid
  modeling of disperse two-phase flows.}},
\newblock \bibinfo{journal}{J. Comput. Phys.} \bibinfo{volume}{246}
  (\bibinfo{year}{2013}) \bibinfo{pages}{145--164}.
\bibitem[{Klar et~al.(2000)Klar, Neunzert, and Struckmeier}]{klar-et-al:2000}
\bibinfo{author}{A.~Klar}, \bibinfo{author}{H.~Neunzert},
  \bibinfo{author}{J.~Struckmeier},
\newblock \bibinfo{title}{{Transition from Kinetic theory to macroscopic fluid
  equations: A problem for domain decompo sition and a source for new
  algorithms}},
\newblock \bibinfo{journal}{Transport Theor. Stat.} \bibinfo{volume}{29}
  (\bibinfo{year}{2000}) \bibinfo{pages}{93--106}.
\bibitem[{Tiwari and Klar(1998)}]{tiwari-klar:1998}
\bibinfo{author}{S.~Tiwari}, \bibinfo{author}{A.~Klar},
\newblock \bibinfo{title}{{An adaptive domain decomposition procedure for
  Boltzmann and Euler equations}},
\newblock \bibinfo{journal}{J. Comput. Appl. Math.} \bibinfo{volume}{90}
  (\bibinfo{year}{1998}) \bibinfo{pages}{223--237}.
\bibitem[{Tiwari et~al.(2013)Tiwari, Klar, Hardt, and
  Donkov}]{tiwari-et-al:2013}
\bibinfo{author}{S.~Tiwari}, \bibinfo{author}{A.~Klar},
  \bibinfo{author}{S.~Hardt}, \bibinfo{author}{A.~Donkov},
\newblock \bibinfo{title}{{Coupled solution of the Boltzmann and Navier-Stokes
  equations in gas-liquid two phase flow}},
\newblock \bibinfo{journal}{Comput. Fluids} \bibinfo{volume}{71}
  (\bibinfo{year}{2013}) \bibinfo{pages}{283--296}.
\bibitem[{Le~Tallec and Mallinger(1997)}]{tallec-mallinger:1997}
\bibinfo{author}{P.~Le~Tallec}, \bibinfo{author}{F.~Mallinger},
\newblock \bibinfo{title}{{Coupling Boltzmann and Navier-Stokes Equations by
  Half Fluxes}},
\newblock \bibinfo{journal}{J. Comput. Phys.} \bibinfo{volume}{136}
  (\bibinfo{year}{1997}) \bibinfo{pages}{51--67}.
\bibitem[{Sugiyama and Kusano(2007)}]{sugiyama-kusano:2007}
\bibinfo{author}{T.~Sugiyama}, \bibinfo{author}{K.~Kusano},
\newblock \bibinfo{title}{{Multi-scale plasma simulation by the interlocking of
  magnetohydrodynamic model and particle-in-cell kinetic model}},
\newblock \bibinfo{journal}{J. Comput. Phys.} \bibinfo{volume}{227}
  (\bibinfo{year}{2007}) \bibinfo{pages}{1340--1352}.
\bibitem[{Daldorff et~al.(2014)Daldorff, T{\'o}th, Gombosi, Lapenta, Amaya,
  Markidis, and Brackbill}]{daldorff-et-al:2014}
\bibinfo{author}{L.~K.~S. Daldorff}, \bibinfo{author}{G.~T{\'o}th},
  \bibinfo{author}{T.~I. Gombosi}, \bibinfo{author}{G.~Lapenta},
  \bibinfo{author}{J.~Amaya}, \bibinfo{author}{S.~Markidis},
  \bibinfo{author}{J.~U. Brackbill},
\newblock \bibinfo{title}{{Two-way coupling of a global Hall
  magnetohydrodynamics model with a local implicit particle-in-cell model}},
\newblock \bibinfo{journal}{J. Comput. Phys.} \bibinfo{volume}{268}
  (\bibinfo{year}{2014}) \bibinfo{pages}{236--254}.
\bibitem[{Markidis et~al.(2014)Markidis, Henri, Lapenta, R{\"o}nnmark, Hamrin,
  Meliani, and Laure}]{markidis-et-al:2014}
\bibinfo{author}{S.~Markidis}, \bibinfo{author}{P.~Henri},
  \bibinfo{author}{G.~Lapenta}, \bibinfo{author}{K.~R{\"o}nnmark},
  \bibinfo{author}{M.~Hamrin}, \bibinfo{author}{Z.~Meliani},
  \bibinfo{author}{E.~Laure},
\newblock \bibinfo{title}{{The Fluid-Kinetic Particle-in-Cell method for plasma
  simulations}},
\newblock \bibinfo{journal}{J. Comput. Phys.}  (\bibinfo{year}{2014})
  \bibinfo{pages}{415--429}.
\bibitem[{Wang et~al.(2015)Wang, Hakim, Bhattacharjee, and
  Germaschewski}]{wang-et-al:2015}
\bibinfo{author}{L.~Wang}, \bibinfo{author}{A.~H. Hakim},
  \bibinfo{author}{A.~Bhattacharjee}, \bibinfo{author}{K.~Germaschewski},
\newblock \bibinfo{title}{{Comparison of multi-fluid moment models with
  particle-in-cell simulations of collisionless magnetic reconnection}},
\newblock \bibinfo{journal}{Physics of Plasmas} \bibinfo{volume}{22}
  (\bibinfo{year}{2015}) \bibinfo{pages}{012108}.
\bibitem[{Johnson(2013)}]{johnson:2013}
\bibinfo{author}{E.~A. Johnson}, \bibinfo{title}{Gaussian-moment relaxation
  closures for verifiable numerical simulation of fast magnetic reconnection in
  plasma}, \bibinfo{type}{Ph.d. thesis}, University of Wisconsin, Madison,
  \bibinfo{year}{2013}. \bibinfo{note}{Eprint arXiv:1409.6985
  [physics.plasm-ph]}.
\bibitem[{Schmitz and Grauer(2006{\natexlab{a}})}]{schmitz1}
\bibinfo{author}{H.~Schmitz}, \bibinfo{author}{R.~Grauer},
\newblock \bibinfo{title}{{Comparison of time splitting and backsubstitution
  methods for integrating {V}lasov's equation with magnetic fields}},
\newblock \bibinfo{journal}{Comp. Phys. Commun.} \bibinfo{volume}{175}
  (\bibinfo{year}{2006}{\natexlab{a}}) \bibinfo{pages}{86--92}.
\bibitem[{Schmitz and Grauer(2006{\natexlab{b}})}]{schmitz2}
\bibinfo{author}{H.~Schmitz}, \bibinfo{author}{R.~Grauer},
\newblock \bibinfo{title}{{Darwin–Vlasov simulations of magnetised plasmas}},
\newblock \bibinfo{journal}{J. Comput. Phys.} \bibinfo{volume}{214}
  (\bibinfo{year}{2006}{\natexlab{b}}) \bibinfo{pages}{738--756}.
\bibitem[{Umeda and Fukazawa(2015)}]{umeda-fukazawa}
\bibinfo{author}{T.~Umeda}, \bibinfo{author}{K.~Fukazawa},
\newblock \bibinfo{title}{{A high-resolution global Vlasov simulation of a
  small dielectric body with a weak intrinsic magnetic field on the K
  computer}},
\newblock \bibinfo{journal}{Earth, Planets and Space} \bibinfo{volume}{67}
  (\bibinfo{year}{2015}) \bibinfo{pages}{49}.
\bibitem[{Filbet et~al.(2001)Filbet, Sonnendr{\"u}cker, and
  Bertrand}]{filbet2001conservative}
\bibinfo{author}{F.~Filbet}, \bibinfo{author}{E.~Sonnendr{\"u}cker},
  \bibinfo{author}{P.~Bertrand},
\newblock \bibinfo{title}{Conservative numerical schemes for the {V}lasov
  equation},
\newblock \bibinfo{journal}{J. Comput. Phys.} \bibinfo{volume}{172}
  (\bibinfo{year}{2001}) \bibinfo{pages}{166--187}.
\bibitem[{Kurganov and Levy(2000)}]{kur2000}
\bibinfo{author}{A.~Kurganov}, \bibinfo{author}{D.~Levy},
\newblock \bibinfo{title}{{A Third-Order Semidiscrete Central Scheme for
  Conservation Laws and Convection-Diffusion Equations}},
\newblock \bibinfo{journal}{SIAM J. Sci. Comput.} \bibinfo{volume}{22}
  (\bibinfo{year}{2000}) \bibinfo{pages}{1461--1488}.
\bibitem[{Shu and Osher(1988)}]{shu88}
\bibinfo{author}{C.~W. Shu}, \bibinfo{author}{S.~Osher},
\newblock \bibinfo{title}{{Efficient Implementation of Essentially
  Non-oscillatory Shock-Capturing Schemes}},
\newblock \bibinfo{journal}{J. Comput. Phys.} \bibinfo{volume}{77}
  (\bibinfo{year}{1988}) \bibinfo{pages}{439--471}.
\bibitem[{Yee(1966)}]{yee1966}
\bibinfo{author}{K.~S. Yee},
\newblock \bibinfo{title}{{Numerical Solution of Initial Boundary Value
  Problems Involving Maxwell's Equations in Isotropic Media}},
\newblock \bibinfo{journal}{IEEE Trans. Antennas Propag.} \bibinfo{volume}{14}
  (\bibinfo{year}{1966}) \bibinfo{pages}{302--307}.
\bibitem[{Taflove and Brodwin(1975)}]{taf1975}
\bibinfo{author}{A.~Taflove}, \bibinfo{author}{M.~E. Brodwin},
\newblock \bibinfo{title}{{Numerical Solution of Steady-State Electromagnetic
  Scattering Problems Using the Time-Dependent Maxwell's Equations}},
\newblock \bibinfo{journal}{IEEE Trans. Microwave Theory Tech.}
  \bibinfo{volume}{23} (\bibinfo{year}{1975}) \bibinfo{pages}{623--630}.
\bibitem[{Tikhonov(1963)}]{tikhonov:1963}
\bibinfo{author}{A.~Tikhonov},
\newblock \bibinfo{title}{{Solution of incorrectly formulated problems and the
  regularization method}},
\newblock \bibinfo{journal}{Soviet Mathematics Doklady} \bibinfo{volume}{4}
  (\bibinfo{year}{1963}) \bibinfo{pages}{1035--1038}.
\bibitem[{Cauchy(1847)}]{cauchy:1847}
\bibinfo{author}{A.~Cauchy},
\newblock \bibinfo{title}{{M{\'e}thode g{\'e}n{\'e}rale pour la r{\'e}solution
  des systemes d’{\'e}quations simultan{\'e}es}},
\newblock \bibinfo{journal}{Comptes Rendus de l'Acad\'{e}mie des Sciences}
  \bibinfo{volume}{25} (\bibinfo{year}{1847}) \bibinfo{pages}{536--538}.
\bibitem[{Press et~al.(2007)Press, Teukolsky, Vetterling, and
  Flannery}]{press:2007}
\bibinfo{author}{W.~H. Press}, \bibinfo{author}{S.~A. Teukolsky},
  \bibinfo{author}{W.~T. Vetterling}, \bibinfo{author}{B.~P. Flannery},
  \bibinfo{title}{{Numerical Recipes 3rd Edition: The Art of Scientific
  Computing}}, \bibinfo{edition}{third} ed., \bibinfo{publisher}{Cambridge
  University Press}, \bibinfo{address}{New York, NY, USA},
  \bibinfo{year}{2007}.
\bibitem[{Leslie and Purser(1995)}]{leslie-purser:1995}
\bibinfo{author}{L.~M. Leslie}, \bibinfo{author}{R.~J. Purser},
\newblock \bibinfo{title}{{Three-dimensional mass-conserving semi-Lagrangian
  scheme employing forward trajectories}},
\newblock \bibinfo{journal}{Monthly Weather Review} \bibinfo{volume}{123}
  (\bibinfo{year}{1995}) \bibinfo{pages}{2551--2566}.
\bibitem[{Birn et~al.(2001)Birn, Drake, and Shay}]{bir2001}
\bibinfo{author}{J.~Birn}, \bibinfo{author}{J.~F. Drake},
  \bibinfo{author}{M.~A. Shay},
\newblock \bibinfo{title}{{Geospace Environmental Modeling (GEM) Magnetic
  Reconnection Challenge}},
\newblock \bibinfo{journal}{J. Geophys. Res.} \bibinfo{volume}{106}
  (\bibinfo{year}{2001}) \bibinfo{pages}{3715--3719}.
\bibitem[{Harris(1962)}]{har1962}
\bibinfo{author}{E.~G. Harris},
\newblock \bibinfo{title}{On a plasma sheath separating regions of oppositely
  directed magnetic field},
\newblock \bibinfo{journal}{Il Nuovo Cimento} \bibinfo{volume}{23}
  (\bibinfo{year}{1962}) \bibinfo{pages}{115--121}.

\end{thebibliography}

\end{document}